\documentclass[12pt]{iopart}
\pdfoutput=1
\eqnobysec
\usepackage{enumerate}
\usepackage{amssymb}
\usepackage{braket}
\usepackage{graphicx}
\usepackage{stmaryrd}
\usepackage[usenames,dvipsnames]{color}
\usepackage[colorlinks,bookmarks=false,citecolor=NavyBlue,linkcolor=OliveGreen,urlcolor=blue]{hyperref}
\newcommand{\be}{\begin{equation}}
\newcommand{\ee}{\end{equation}}
\newcommand{\bea}{\begin{eqnarray}}
\newcommand{\eea}{\end{eqnarray}}
\newcommand{\ba}{\begin{aligned}}
\newcommand{\ea}{\end{aligned}}
\newcommand{\1}{{\rm I}}
\newcommand{\nn}{\nonumber\\}

\begin{document}
\title[Quantum XY model with OBC]{Local conservation laws in spin-$\frac{1}{2}$ XY chains with open boundary conditions}
\author{Maurizio Fagotti}
\address{D\'epartement de Physique, \'Ecole Normale Sup\'erieure/PSL Research University, CNRS, 24 rue Lhomond, 75005 Paris, France}
\begin{abstract}
We revisit the conserved quantities of the spin-$\frac{1}{2}$ XY model with open boundary conditions. 
In the absence of a transverse field, we find new families of local charges and  show that half of the seeming conservation laws are conserved only if the number of sites is odd.  
In even chains the set of noninteracting charges is abelian, like in the periodic  case when the number of sites is odd. In odd chains the set is doubled and becomes non-abelian, like in even periodic chains.  The dependence of the charges on the parity of the chain's size undermines the common belief that the thermodynamic limit of diagonal ensembles exists.
We consider also the transverse-field Ising chain, where the situation is more ordinary. The generalization to the XY model in a transverse field is not straightforward and we propose a general framework to carry out similar calculations. We conjecture the form of the bulk part of the local charges and discuss the emergence of quasilocal conserved quantities. We provide evidence that in a region of the parameter space there is a reduction of the number of quasilocal conservation laws invariant under chain inversion. 
As a by-product, we study a class of block-Toeplitz-plus-Hankel operators 
and identify the conditions that their symbols satisfy in order to commute with a given block-Toeplitz.
\end{abstract}
\maketitle
\tableofcontents
\title[Quantum XY model with OBC]{}
\section{Introduction and the basics}
The local and quasilocal conservation laws of a quantum many-body system play a key role in the late time dynamics of local observables after a so-called global quench~\cite{BMD:XY,Greiner, SPS:2004, CC:quench,cradle} (see also \cite{E:rev} and references therein). This term is usually used to refer to the non-equilibrium time evolution of the ground state of a local Hamiltonian after an abrupt change of a global Hamiltonian parameter. Quite generally in the thermodynamic limit the reduced density matrices of finite subsystems approach stationary values.
In the last ten years it has been realized~\cite{Rigol,CE:climit,FE:rdm,clustdec,D:stat} that the stationary behavior  
can be described by effective stationary states determined only by a tiny subset of the integrals of motion. In particular, it was shown that local integrals of motion encode information that is not generally replaceable but becomes less and less important the larger the range of the conservation law is~\cite{FE:rdm}\footnote{Since in interacting models like the XXZ spin-$\frac{1}{2}$ chain the set of the local conserved quantities does not seem to be complete, this picture has been also extended to quasilocal conservation laws with a finite typical range~\cite{ql}.}. 
Consequently, the stationary properties of local observables distinguish between generic and integrable models, having the latter infinitely many local conservation laws, instead of the few (if not only the Hamiltonian) local conserved quantities that characterize generic models.  

This situation results in instabilities in the stationary behavior of local observables after global quenches in integrable models, where generic perturbations produce global changes.  
With weak integrability breaking, dynamics occur on different timescales and one can identify slow transitions, prethermalization~\cite{preT0,preT1,preTS, preTNeil, preT2, preTBruno,preTDemler, FC:pretSvN} or pre-relaxation~\cite{F:super,BF:mf}, between globally different (quasi-stationary) states.

In \cite{F:lgQ} it was pointed out that this situation is not specific to global perturbations. 
Any perturbation that spoils (or augments) a set of local conservation laws has  what it takes to activate transitions in non-equilibrium states.
In particular, spatially localized perturbations can drive crossovers between globally different states and, in turn, some global properties of the state can be controlled by acting on a local part of the system. 

Changing the boundary conditions from periodic (PBC) to open (OBC) is probably the minimal local perturbation that is known~\cite{GM:open,QFT:open} to affect the set of the local conservation laws in integrable models. 
A subset of the local conservation laws with PBC are spoiled by the boundaries while the others are only deformed close to the edges, remining undistinguishable in the bulk from the corresponding charges with PBC. In addition, there can be  quasilocal charges that zero in the bulk (edge modes). 

From the perspective of \cite{F:lgQ}, the basic question to answer is which are the local conservation laws that survive the local defect, \emph{i.e.} the boundary. It is indeed reasonable to expect that these will be the quantities that characterize the stationary properties of finite subsystems at infinitely large times after the quench. 

Similarly to what happens in translation invariant systems~\cite{GM:periodic}, also in the presence of boundaries the local charges can be formally obtained in the framework of algebraic Bethe ansatz~\cite{Korepin} as the derivatives of the logarithm of an appropriate transfer matrix~\cite{TMboundary}. This was used in \cite{GM:open} to deduce some general properties of the local conservation laws, like invariance under chain inversion. 
However, to the best of our knowledge, no necessary and sufficient condition for a conserved quantity in the bulk has been established that guarantees the charge to ``survive'' the boundary in a spin chain model.
For example, taking into account the results of Ref.~\cite{GM:open} for the quantum XY model in a transverse field, chain inversion does not seem to be a sufficient condition for the presence of a local charge. In fact, we will show that it is not even necessary. 
 
The results of  \cite{GM:open, GM:periodic} have been already revisited~\cite{PP:ql,ql} to take into account that the standard representation of the transfer matrix is not generally sufficient to obtain all the (quasi)local conservation laws.  
There are however also other reasons to reconsider the problem of the conserved quantities in the quantum XY model in particular. First of all, Ref.~\cite{F:super} exhibited new families of local charges in the XY model with PBC, which could have counterparts in the OBC case. 
Second, while in interacting integrable models the algebraic Bethe ansatz provides an elegant framework to investigate the local conservation laws, surprisingly an analogous workbench in models that can be mapped to noninteracting fermions with OBC has not yet been established (instead, in translation invariant chains, convenient formalisms have been already proposed~\cite{F:super, FE:rdm, clIsing}). As a result, the brute force search of Ref.~\cite{GM:open} for the XY model appears less refined and systematic than the transfer matrix formalism for interacting models.

The main goal of this paper is to provide a framework where to address the construction of the local conservation laws in models with OBC that can be mapped to noninteracting fermions but for which the Bethe ansatz formalism is not always effective. In particular, we will study the local charges of the spin-$\frac{1}{2}$ XY model with open boundary conditions, both in the finite and in the semi-infinite chain.

We use two different but somehow complementary approaches: a direct method that relies on exact diagonalization, and a more abstract procedure that is based on a correspondence between quadratic forms of noninteracting fermions and Toeplitz+Hankel operators. 
Specifically, we show that the research of local charges can be reduced to the construction of a particular class of block-Toeplitz-plus-Hankel operators that commute with a given block-Toeplitz. 
The  latter problem can be set in a rather elegant way and we identify the conditions that are satisfied by the so-called symbols of the block-Toeplitz-plus-Hankel operators of the local conserved quantities.

Our analysis reveals some subtleties  that were originally overlooked. 
In particular, we show that conservation laws with range odd are present only if the chain's length is odd. 
We also find new families of local conservation laws that break one-site shift invariance in the bulk and even a family that is not invariant under chain inversion.
 
The final part of the paper is about the effects of a nonzero transverse field. 
While this is little complication in the chain with PBC, it makes the problem much more involved in the OBC case. We show that in the transverse-field Ising chain the construction remains simple and chain inversion turns out to be a sufficient criterion to identify all the local charges. On the other hand, the result is not easily generalizable to the XY model in a transverse field, where one is forced to consider also the emergence of quasilocal conserved quantities. In that case, we propose a conjecture about the form of the bulk part of the local conservation laws and identify a curve in the parameter space that separates a region with an Ising-like set of (quasi)local charges from a region where there are less conserved quantities which are invariant under chain inversion.

\subsection{Locality and quasilocality}\label{s:def}
The terminology that is commonly used to refer to the local properties of operators acting on spin chains can be confusiong. We report here the definitions that will be used in this paper. 

We say that an operator $\mathcal O$ is  `localized' in a connected finite subsystem $S$ if
\be\label{eq:loc}
\mathcal O=\Tr_{\bar S}[\mathcal O]\otimes \frac{\1_{\bar S}}{\Tr_{\bar S}[\1]}\qquad {\rm (localized)}\, ,
\ee
where $\bar S$ is the complement of  $S$.
The range of $\mathcal O$ is the length of the smallest subsystem for which the equality holds. 
Clearly, the class of localized operators is closed under commutation. 

The standard extension of the class \eref{eq:loc} is given by the operators that can be approximated by localized operators up to exponentially small corrections in the length of the reference subsystem.
Formally, one could define such operators as follows. Given an infinite sequence of subsystems $S_n\subset S_{n+1}$ with increasing length $|S_n|$ such that $\lim_{n\rightarrow\infty}S_n=\Omega$ ($\Omega$ is the entire chain), we consider operators $\mathcal O$ such that 
\be\label{eq:quasiloc}
\forall |S|>\xi,\quad \frac{\parallel\mathcal O-\Tr_{\bar S}[\mathcal O]\otimes \frac{\1_{\bar S}}{\Tr_{\bar S}[\1]}\parallel}{\parallel\mathcal O\parallel}\leq e^{-|S|/\xi}\qquad {\rm (\mbox{quasilocalized})}\, ,
\ee
where $\parallel\cdot\parallel$ is the operator norm (the maximal eigenvalue, in absolute value) and $\xi$ is a finite length.  We qualify these as `quasilocalized' operators and call `typical range' the minimal length $\xi$ for which it is possible to find a sequence $S_n$ such that \eref{eq:quasiloc} holds.

We say that $Q$ is \emph{local} if, for any localized operator $\mathcal O$, the commutator $[Q,\mathcal O]$ is localized as well.  Using this definition, a localized operator is also local. 
Analogously, we call it \emph{quasilocal} if the commutator with any quasilocalized operator is quasilocalized. Clearly, a local operator is also quasilocal. 
Roughly speaking, a (quasi)local operator is a linear combination of (quasi)localized operators. 
We warn the reader  that in the scientific literature there are alternative (weaker) definitions of quasilocality based on the Hilbert-Schmidt norm~\cite{PP:ql}. 

Let $Q_i$ be conserved operators $[H,Q_i]=0$.
Using the Jacobi's identity one can immediately show that the commutator of two (quasi)local conservation laws is conserved and \mbox{(quasi)local}, being its commutator with any \mbox{(quasi)localized} operator a combination of two nested commutators with the \mbox{(quasi)local} charges:
\begin{equation*}
[[Q_1,Q_2],\mathcal O]=[Q_1,[Q_2,\mathcal O]]-[Q_2,[Q_1,\mathcal O]]\, .
\end{equation*}

\subsection{The model}
We consider the Hamiltonian of the spin-$\frac{1}{2}$ XY chain in a transverse field
\be\label{eq:H}
H=-J \sum_{j=1}^{L-1}\Bigl(\frac{1+\gamma}{4}\sigma_\ell^x\sigma_{\ell+1}^x+\frac{1-\gamma}{4}\sigma_\ell^y\sigma_{\ell+1}^y\Bigr)-\frac{J h}{2}\sum_{j=1}^L\sigma_\ell^z
\ee
where $\sigma_\ell^\alpha$ act like Pauli matrices on site $\ell$ and like the identity elsewhere; the constant $J$ is irrelevant to our purposes and will be chosen so as to ease the notations. 

The model is exactly solvable~\cite{LSM:XY} and, in its parameter space, includes the transverse-field Ising chain (TFIC, $|\gamma|=1$), which is a crucial paradigm for quantum critical behavior, and the XX model ($\gamma=0$), which corresponds to the noninteracting limit of the XXZ Heisenberg spin-$\frac{1}{2}$ chain. 

In the first part of the paper we will consider the case $h=0$, which will be simply referred to  as ``XY model''. The effects of a non-vanishing transverse field will be addressed in the second part.

\subsubsection{Symmetries.}\label{ss:sym}
The XY model in a transverse field \eref{eq:H} is invariant under spin flip about $z$, realized by the operator
\begin{equation}\label{eq:Piz}
\Pi^{z}=\prod_{j=1}^L\sigma_j^z\, .
\end{equation}
It is also invariant under chain inversion 
\begin{equation*}
{\bf R}: \sigma_\ell^\alpha\rightarrow \sigma_{L+1-\ell}^\alpha\, ,
\end{equation*} 
with $\alpha\in\{x,y,z\}$.
In addition, the transformation $\prod_{j=1}^{\lfloor (L+1)/2\rfloor}\sigma_{2j-1}^x\prod_{j=1}^{\lfloor L/2\rfloor}\sigma_{2j}^y$\footnote{We use the symbol $\lfloor m \rfloor$ to indicate the largest integer smaller than or equal to $m$.}
maps the Hamiltonian in minus itself, and hence maps the set of the charges in itself. 

Without loss of generality, we can assume $\gamma\geq 0$: the sign of $\gamma$ can be changed by a rotation about $z$. Analogously, we can assume $h\geq 0$, indeed $\Pi^x=\prod_{j=1}^L \sigma_j^x$ reverses the sign of $h$. 
\paragraph{XY model ($h=0$).}
In the absence of a transverse field, $\gamma$ can  be chosen to be smaller than $1$.
This is related to the transformation  $V^x=\prod_{j=1}^{\lfloor L/2 \rfloor}\sigma^x_{2n}$: $V^x$ changes the sign of the coupling constant in the $y$ direction, which results in the same Hamiltonian, with the parameters transformed as $J\rightarrow J\gamma$ and $\gamma\rightarrow \gamma^{-1}$
\begin{equation*}
\fl\quad  -J \sum_{j=1}^{L-1}\frac{1+\gamma}{4}\sigma_\ell^x\sigma_{\ell+1}^x+\frac{\gamma-1}{4}\sigma_\ell^y\sigma_{\ell+1}^y=-(J\gamma) \sum_{j=1}^{L-1}\frac{1+\gamma^{-1}}{4}\sigma_\ell^x\sigma_{\ell+1}^x+\frac{1-\gamma^{-1}}{4}\sigma_\ell^y\sigma_{\ell+1}^y\, .
\end{equation*}
Thus, we will generally assume $0<\gamma< 1$ (the case $\gamma=1$ is not very interesting since it corresponds  to the classical Ising model). 
Besides spin flip about $z$, the Hamiltonian is also invariant under spin flip about $x$ and $y$,  realized by the operators
\begin{equation*}
\Pi^{x(y)}=\prod_{j=1}^L\sigma_j^{x(y)}\, .
\end{equation*}

\subsubsection{Fermion representation.}\label{ss:frep}
The Hamiltonian \eref{eq:H} can be mapped to a quadratic form of fermions by a Jordan-Wigner transformation. In particular we can write
\be\label{eq:Hquad}
H=\frac{1}{4}\sum_{j,\ell=1}^{2L}a_j \mathcal H_{j \ell}a_\ell\, .
\ee
where $a_j$ are the Majorana fermions 
\be\label{eq:a}
a_{2\ell-1}=a_\ell^x=\prod_{j=1}^{\ell-1}\sigma_j^z \sigma_{\ell}^x\qquad a_{2\ell}=a_\ell^y=\prod_{j=1}^{\ell-1}\sigma_j^z \sigma_{\ell}^y\, ,
\ee
satisfying the algebra $\{a_\ell,a_n\}=2\delta_{\ell n}$.
Under chain inversion and spin flip the Majorana fermions transform as follows:
\begin{eqnarray*}
{\bf R}a_\ell {\bf R} =  (-1)^{\ell+1}i \Pi^z a_{2L+1-\ell}\\
\Pi^x a_\ell\Pi^x=(-1)^{\lceil\frac{\ell-1}{2}\rceil} a_\ell\\
\Pi^y a_\ell\Pi^y=(-1)^{\lceil\frac{\ell+1}{2}\rceil} a_\ell\\
\Pi^z a_\ell\Pi^z=-a_{\ell}\, .
\end{eqnarray*}
The matrix $\mathcal H$ is Hermitian and purely imaginary block-Toeplitz with $2$-by-$2$ blocks 
\be\label{eq:blockT}
\fl\qquad  \left(
\begin{array}{cc}
\mathcal H_{2\ell-1, 2n-1}&\mathcal H_{2\ell-1, 2n}\\
\mathcal H_{2\ell, 2n-1}&\mathcal H_{2\ell, 2n}
\end{array}\right)=
J\Bigl(\delta_{n,\ell+1}\frac{\sigma^y+i\gamma \sigma^x}{2}+ \delta_{\ell,n+1} \frac{\sigma^y-i\gamma\sigma^x}{2}-h\sigma^y  \delta_{\ell n} \Bigr)\, .
\ee
The eigenstates and eigenvalues of $H$ \eref{eq:H} can be obtained from the eigenvalues and eigenvectors of $\mathcal H$ \eref{eq:Hquad} as follows. Being $\mathcal H$ a purely imaginary skew-symmetric matrix, the eigenvalues  come in pairs with opposite sign and an orthogonal transformation $\mathcal R$ brings the matrix to block-diagonal form
\begin{equation*}
\mathcal H=\mathcal R \mathcal E\otimes \sigma^y \mathcal R^t\, .
\end{equation*}
Here $\mathcal E$ is a diagonal matrix with nonnegative diagonal elements $\varepsilon_i$. As a consequence, a unitary transformation that diagonalizes $\mathcal H$ ($\mathcal H=\mathcal U  \mathcal E\otimes \sigma^z \mathcal U^\dag$) is given by
\begin{equation*}
\left(
\begin{array}{c}
\mathcal U_{j, 2i-1}\\
\mathcal U_{j, 2i}
\end{array}
\right)=\frac{1}{\sqrt{2}}[e^{i\frac{\pi}{4}}\1+e^{-i\frac{\pi}{4}} \sigma^x]
\left(
\begin{array}{c}
\mathcal R_{j, 2i-1}\\
\mathcal R_{j, 2i}
\end{array}
\right)\, .
\end{equation*}
From this equation it follows that  $\mathcal U_{j, 2i}=\mathcal U_{j, 2i-1}^* $. Using that $\mathcal U$ is unitary one can check that  the operators
\be\label{eq:bU}
b^\dag_i=\frac{1}{\sqrt{2}}\sum_j a_j \mathcal U_{j, 2i-1} \qquad b_i=\frac{1}{\sqrt{2}}\sum_j a_j \mathcal U_{j, 2i}
\ee
satisfy the fermion anticommutation relations
\begin{equation*}
\{b^\dag_i,b_j\}=\delta_{i j}\qquad \{b^\dag_i,b^\dag_j\}=\{b_i,b_j\}=0\, .
\end{equation*}
In addition, the commutator with the Hamiltonian reads
\begin{equation*}
\fl\qquad [H,b^\dag_i]=\frac{1}{4\sqrt{2}}\sum_{j,\ell,n=1}^{L}\mathcal H_{j \ell} \mathcal U_{n, 2i-1} [a_j a_\ell,a_n]=\frac{1}{\sqrt{2}}\sum_j a_j [ \mathcal U \mathcal E\otimes\sigma^z]_{j, 2i-1}=\varepsilon_i b^\dag_i\, .
\end{equation*}
Observing that the Hamiltonian is traceless, this implies 
\begin{equation}\label{eq:Hdiag}
H=\sum_i\varepsilon_i\Bigl(b^\dag_i b_i-\frac{1}{2}\Bigr)\, .
\end{equation}
In other words, the nonnegative eigenvalues of $\mathcal H$ are the excitation energies of $H$ and the fermions that diagonalize $H$ can be reconstructed from the eigenvectors of $\mathcal H$ by means of \eref{eq:bU}. 
Thus, one can focus on the eigenvalue problem 
\be\label{eq:eig_prob}
\mathcal H \vec{\mathcal U}_i=\varepsilon_i  \vec{\mathcal U}_i\, ,
\ee
with $[\vec{\mathcal U}_i]_j=\mathcal U_{j i}$.  

The representation \eref{eq:Hquad} is also useful to address the problem of the local conserved quantities. One can easily show
\begin{equation*}
\Bigl[\frac{1}{4}\sum_{j,\ell=1}^{2L}(a_j \mathcal A_{j \ell}a_\ell),\frac{1}{4}\sum_{j,\ell=1}^{2L}(a_j \mathcal B_{j \ell}a_\ell)\Bigr]=\frac{1}{4}\sum_{j,\ell=1}^{2L}a_j [\mathcal A,\mathcal B]_{j \ell}a_\ell\, ,
\end{equation*}
which implies that the quadratic forms of fermions $Q$ that commute with $H$ must be characterized by matrices $\mathcal Q$ that commute with $\mathcal H$
\begin{equation*}
[H,Q]=0\Rightarrow [\mathcal H,\mathcal Q]=0\, .
\end{equation*}
Locality is equivalent to ask for $\mathcal Q$ having only a finite number of diagonals different from zero (around the main diagonal); analogously, quasilocality means that the elements decay exponentially fast with the difference between row and column indices. We warn the reader that the local conservation laws do not need to take the form \eref{eq:Hdiag} when the dispersion relation is degenerate ($\varepsilon_i=\varepsilon_j$ for some $i\neq j$) \cite{F:super}. In addition, generally there can be also (quasi)local conservation laws which are not quadratic in the fermions, but this seems to happen only when the dispersion relation has exceptional symmetries. 
  
In conclusion, the mapping $Q\rightarrow\mathcal Q$ allows to reduce  the complexity of the problem exponentially (in the system size). However, we point out that without any further simplification the problem would be still unsolvable.

\subsection*{Organization of the paper}
The rest of the paper is organized as follows.
In \Sref{s:summary} we present our main results for the semi-infinite chain. 
Sections \ref{s:excXY} and \ref{s:lclXY} are focussed on the XY model without transverse field: the excitations are reported in \sref{s:excXY} and the local conservation laws are constructed in \sref{s:lclXY}.   
The reader who is not interested in the details of this exceptional case can  jump directly to \sref{s:TH}. 
\Sref{s:TH} introduces a general formalism to determine the local conservation laws of quadratic Hamiltonians with open boundary conditions.
The approach is explicitly applied to the transverse-field Ising chain.  We also report a preliminary study of the conservation laws of the quantum XY model in a transverse field and conjecture the form of their bulk part.  \Sref{s:conc} collects some conclusive remarks. 
The main text is followed by two appendices. 
In \ref{a:TH} there is a proof of the conditions that block-Toeplit-plus-Hankel operators satisfy in order to commute with a block-Toeplitz. 
In \ref{A:direct} we report some details of the diagonalization of the model.

\section{Summary and discussion of the results}\label{s:summary}
In this section we report the expressions of the local and quasilocal conservation laws in the semi-infinite XY chain with $h=0$ or $\gamma=1$ and a conjecture about the form of the bulk part of the local charges for generic $h$ and $\gamma$. We also present a mathematical framework that allows one to search for the local charges without relying on the exact diagonalization of the model.

\subsection{Local conservation laws}\label{s:lcl}
\begin{table}
\begin{center}
\begin{tabular}{c||c|c|c}
Model&Family&Toeplitz symbol $\hat q_T(e^{ik})$&Hankel symbol $\hat q_H(e^{ik})$\\
\hline\hline
XY{\scriptsize($h=0$)}\\
\hline
&$I_j^{+_{\rm bond}(e)}$&$\cos(j k)\varepsilon_k\sigma^x e^{-i\frac{k}{2}\sigma^z}\otimes \sigma^y e^{-i\theta_k\sigma^z}$&$\frac{e^{i(j-\frac{1}{2})k}}{2}\varepsilon_k(\sigma^x\otimes\sigma^y)e^{i\theta_k\sigma^z\otimes\sigma^z}$\\
&$J_j^{+_{\rm bond}(e)}$&$\cos(j k)\varepsilon_k\sigma^y e^{-i\frac{k}{2}\sigma^z}\otimes \sigma^x e^{-i\theta_k\sigma^z}$&$\frac{e^{i(j-\frac{1}{2})k}}{2}\varepsilon_k(\sigma^y\otimes\sigma^x)e^{i\theta_k\sigma^z\otimes\sigma^z}$\\
&$W_j^{-_{\rm site}(o)}$&$\cos(j k)\varepsilon_k(\sigma^y e^{-i\frac{k}{2}\sigma^z}\otimes\sigma^z)e^{-i\theta_k\sigma^z\otimes \sigma^z}$&$\frac{e^{i (j-\frac{1}{2})k}}{2}\varepsilon_k\sigma^y\otimes(\sigma^z e^{i\theta_k\sigma^z})$\\
&$W_j^{+_{\rm site}(o)}$&$\cos(jk)\varepsilon_k [\varepsilon_k \1\otimes\sigma^y e^{-i\theta_k\sigma^z}]e^{-i\theta_k\sigma^z\otimes \sigma^z}$&$\frac{e^{i (j-\frac{1}{2})k}}{2}\varepsilon^2_ke^{i\frac{k}{2}\sigma^z} \otimes\sigma^y$\\
{}[XX {\scriptsize($\gamma=0$)}] &$S^z$&$\cos(\frac{k}{2})\varepsilon_k \1\otimes\sigma^y e^{-i\theta_k\sigma^z}$&\\
\hline
TFIC {\scriptsize$(\gamma=1)$}\\
\hline
&$I_j^{+}$&$\cos(jk)[(\cos k-h)\sigma^y+\sin k\sigma^x]$&$\frac{e^{i(j-1)k}}{2}\sigma^y(e^{i k}-h)$
\end{tabular}
\caption{Local conservation laws that are nonzero in the bulk. The dispersion relation is $\varepsilon_k=(\cos^2\frac{k}{2}+\gamma^2\sin^2\frac{k}{2})^{1/2}$ and the Bogoliubov angle $e^{i\theta_k}=(\cos \frac{k}{2}+i\gamma\sin \frac{k}{2})/\varepsilon_k$.}\label{t:charge}
\end{center}
\end{table}

The local conservation laws that are nonzero in the bulk have the form
\be\label{eq:Qrep0}
Q=\frac{1}{4}\sum_{n, m=1}^\infty a_n \mathcal Q_{n m}a_m\, .
\ee
Here $\mathcal Q$ is a block-Toeplitz-plus-Hankel operator with elements
\be\label{eq:QTH}
\fl \qquad[\mathcal Q]_{2\kappa \ell+j, 2\kappa \ell' +j'}=\int_{-\pi}^\pi\frac{\mathrm d k}{2\pi}e^{-i(\ell-\ell')k}[\hat q_{T,\kappa}(e^{i k})]_{j j'}-\int_{-\pi}^\pi\frac{\mathrm d k}{2\pi}e^{-i(\ell'+\ell+1)k}[\hat q_{H,\kappa}(e^{ik})]_{j j'}
\ee
where $0\leq \ell,\ell'<\infty$, $1\leq j,j'\leq 2\kappa$ and $\kappa$ is an arbitrary positive integer that represents the number of the chain sites that are formally associated with a single block. 
The two matrix functions $\hat q_{T,\kappa}(e^{i k})$ and $\hat q_{H,\kappa}(e^{ik})$ are the symbols of the Toeplitz and of the Hankel part, respectively. 
The block-Toeplitz part is reminiscent of the structure of the charges in the infinite chain and the block-Hankel part is used to parametrize the boundary term (as long as the boundary term is localized this ansatz is not at all restrictive).
According to the picture presented at the end of \sref{ss:frep}, the symbols associated with local operators  have a finite number of nonzero Fourier coefficients.  
The explicit values of the symbols for the XY model ($h=0$) and for the transverse-field Ising chain ($\gamma=1$) are reported in \tref{t:charge}\footnote{Ref.~\cite{F:super} defined the phase in the Toeplitz part with the opposite sign (\emph{i.e.} $k\rightarrow -k$).}. 
For $h=0$ the symbols displayed are 4-by-4 matrices ($\kappa=2$), whereas for $\gamma=1$ they are 2-by-2 ($\kappa=1$).
Up to boundary terms, the corresponding conservation laws are two-site shift invariant.
The notations for the families of charges manifest their transformation rules in the bulk: 
\begin{itemize}
\item[-] Letter $I$ ($J$) denotes evenness (oddness) under a one-site shift; letter $W$ is instead associated with charges that do not transform well under a shift by one site. 
\item[-] Attribute $+_{\rm bond/site}$ stands for reflection symmetry about a bond or a site;  $-_{\rm bond/site}$ is instead oddness under the same reflections; if bond/site is not indicated, the property refers to both cases.
\item[-] Attributes $(e)$ and $(o)$ denote evenness and oddness, respectively, under spin flip about x.
\end{itemize}
We point out that the Hamiltonian is given by $H=I_0^{+ (e)}$, for which the symbol of the Hankel part can be redefined to be zero (only the analytic part of $\frac{1}{z}\hat q_{H,\kappa}(z)$ gives a nonzero contribution in \eref{eq:QTH}). 

Let us focus on the bulk properties of the local charges.
The presence of conservation laws ($W$) for which it is not possible to find linear combinations that transform well under a shift by one site is a specific property of the chain with open boundary conditions: unexpectedly, the boundary is sufficient to induce the breaking of one-site shift invariance.
However, we note that in the isotropic limit (the XX model, $\gamma=0$) also the families $W$ have definite parity under a shift by one site: in particular $W_j^{-_{\rm site}(o)}\rightarrow J_j^{-_{\rm site}(o)} $ and $W_j^{+_{\rm site}(o)}\rightarrow I_j^{+_{\rm site}(o)} $. Moreover, for $\gamma=0$  there is an additional charge: the total spin in the $z$ direction (but the  conservation law that is quasilocalized at the boundary is absent, \emph{cf.} \sref{eq:ssq-loc}). 

\begin{table}
\begin{center}
\begin{tabular}{c||c|c}
Model&Extinct family&Toeplitz symbol\\
\hline\hline
XY{\scriptsize($h=0$)}\\
\hline
&$I_j^{-_{\rm bond}(e)}$&$\sin(j k) \1\otimes\1$\\
&$J_j^{-_{\rm bond}(e)}$&$\sin (j k)\sigma^z\otimes\sigma^z$\\
&$\tilde W_j^{-_{\rm site}(o)}$&$\sin((j-\frac{1}{2}) k)\varepsilon_k (\sigma^x e^{-i\frac{k}{2}\sigma^z}\otimes\1)e^{-i\theta_k\sigma^z\otimes \sigma^z}$\\
&$\tilde W_j^{+_{\rm site}(o)}$&$\sin((j-\frac{1}{2})k)\varepsilon_k [\varepsilon_k \sigma^z\otimes\sigma^x e^{-i\theta_k\sigma^z}]e^{-i\theta_k\sigma^z\otimes \sigma^z}$\\
\hline
TFIC {\scriptsize$(\gamma=1)$}\\
\hline
&$I_j^{-}$&$\sin(jk)\1$
\end{tabular}
\caption{Local conservation laws of the periodic chain that become extinct. The notations are the same as in \tref{t:charge}.}\label{t:ext}
\end{center}
\end{table}

In \tref{t:ext} we report the local conservation laws of the chain with periodic boundary conditions that do not have counterparts in the chain with open boundary conditions. 
In a quench setup, the extinct charges represent the global information from the bulk that the boundary is able to destroy (at infinite times).

For the quantum XY model in a transverse field, we have derived the explicit expression of a few of the most local conservation laws which are invariant under chain inversion and inferred from that the form of the bulk part of a generic local conserved quantity which is invariant under chain inversion. Our conjecture for the symbol of the Toeplitz part reads as
\be\label{eq:conj0}
\hat q_{j;T,1}(e^{i k})=\Bigl(\cos k+\frac{h}{\gamma^2-1}\Bigr)^{2j}\hat e_{T,1}(k)\, ,
\ee
where $\hat e_{T,1}(k)$ is the symbol of the Hamiltonian
\be
\hat e_{T,1}(e^{ik})=(\cos k-h)\sigma^y+\gamma \sin k\sigma^x\, .
\ee
In practice this means that the charges with OBC are in relation to the charges with PBC that have the form
\be
Q^{\rm (PBC)}_j=\int_{-\pi}^\pi\frac{\mathrm d k}{2\pi} \Bigl(\cos k+\frac{h}{\gamma^2-1}\Bigr)^{2j}\varepsilon(k)b^\dag_k b^{\phantom \dag}_k\, ,
\ee
where $\varepsilon(k)$ is the dispersion relation and $b^\dag_k$ are the Boboliubov fermions that diagonalize the model $\{b_p,b^\dag_k\}=2\pi \delta(k-p)$. 
We point out that the absence of odd powers of $\cos_q(k)=\cos k+q$, with $q=h/(\gamma^2-1)$, is indicative of the model having a reduced number of local conservation laws (invariant under chain inversion) but does not necessarily imply that the  charges corresponding to odd powers are spoiled by the boundary.
For example, for $h>|\gamma^2-1|$ we find (\sref{s:XY}) that the apparently missing charges are in fact conserved if one allows for a quasilocalized boundary part; in addition, they are not linearly independent of \eref{eq:conj0}. For $h<|\gamma^2-1|$ the  missing charges are instead linearly independent. Although our analytic analysis is not sufficient to rule out the possibility that they might correspond to conservation laws with a quasilocalized contribution at the boundary, in \sref{s:XYhanalysis} we provide some numerical evidence that this should not be the case. 

We point out that the curve $h=|\gamma^2-1|$ was shown~\cite{ProsenXY} to correspond to a far-from-equilibrium quantum phase transition in the open\footnote{The attribute ``open'' is used here to indicate that the system is coupled with a bath.} XY chain where the density matrix evolves according to a Lindblad master equation and the Lindblad operators act only on the boundaries. 

Although we have not investigated the presence of conservation laws that are odd under chain inversion, we point out that for $h<|\gamma^2-1|$ there are indirect evidences of their presence. 

\subsection{Generalization: a lemma for block-Topelitz-plus-Hankel operators}\label{s:lclTH}

The symbols of block-Toeplitz-plus-Hankel operators \eref{eq:QTH} associated with local Hermitian quadratic forms of fermions  satisfy the following properties (when there is ambiguity, the argument of the functions can be assumed to be real):
\begin{itemize}
\item[-] a finite $n$ exists such that $z^{n} \hat q_{T,\kappa}(z)$ and $\hat q_{H,\kappa}(z)$ are matrix polynomials;
\item[-] $\hat q_{T,\kappa}(z)=-\hat q_{T,\kappa}^\ast (z)=-\hat q_{T,\kappa}^t (1/z)$;
\item[-] $\hat q_{H,\kappa}(z)=-\hat q_{H,\kappa}^\ast(z)=-q_{H,\kappa}^t(z)$.
\end{itemize}
In addition, one can always assume $\hat q_H(0)=0$ (\emph{cf.} \eref{eq:QTH}). 
We note that the range of the operator is proportional to the (maximal) degree of the polynomials and to the block's size $\kappa$. Therefore, increasing $\kappa$ the degree of the polynomials decreases and for $\kappa$ large enough the symbols become linear in $z$ (and in $1/z$, for the Toeplitz part).  
It is convenient to introduce the notation
\begin{equation*}
\hat q_{T,\kappa}(z)=\hat q_{T,\kappa}^+(z)+\hat q_{T,\kappa}^0+\hat q_{T,\kappa}^-(1/z)\, ,
\end{equation*}
where $\hat q_{T,\kappa}^{\pm}(z)$ are polynomials in $z$ with no constant term ($\hat q_{T,\kappa}^{\pm}(0)=0$). 
From the previous relations it follows
\begin{eqnarray*}
{}[\hat q_{T,\kappa}^0]^t=-\hat q_{T,\kappa}^0\\
{}[\hat q_{T,\kappa}^+]^t(z)=-\hat q_{T,\kappa}^- (z)\, .
\end{eqnarray*}
In the following, if $\frac{\mathrm d^2}{\mathrm d z^2}\hat q_{T,\kappa}^\pm (z)=0$, with an abuse of notation we will write $\hat q_{T,\kappa}^\pm (z)=z \hat q_{T,\kappa}^\pm $.

Imposing vanishing commutator with the Toeplitz operator associated with the Hamiltonian, we find that the symbols of the local charges must satisfy the following conditions (see \ref{a:TH}):
\be\label{eq:sys}
\fl\qquad\qquad\left\{
\begin{array}{l}
[\hat e_{T,\kappa}(z),\hat q_{T,\kappa}(z)]=0\\
\frac{\partial^2}{\partial z^2}\Bigl[\hat e_{T,\kappa}(z)\hat q_{H,\kappa}(z)-\hat q_{H,\kappa}(z) \hat e_{T,\kappa}(1/z)\Bigr]=\Bigl[\hat q_{T,\kappa}^{+''}(0) -2\hat q_{H,\kappa}'(0)\Bigr]\hat e_{T,\kappa}^-\\
\frac{\partial^2}{\partial z^2}\Bigl[\frac{\hat q_{T,\kappa}^+(z)}{z} -\hat q_{H,\kappa}(z)\Bigr]\hat e_{T,\kappa}^-=0\\
\Bigl[\hat q_{T,\kappa}^{+''}(0)-2\hat q_{H,\kappa}'(0)\Bigr]\hat e_{T,\kappa}^- {\rm\ is\ antisymmetric}\\
\hat q_{T,\kappa}^{+'}(0)\hat e_{T,\kappa}^- +\frac{1}{2}\hat q^{''}_{H,\kappa}(0) \hat e_{T,\kappa}^++\hat q'_{H,\kappa}(0)\hat e_{T,\kappa}^0 {\rm\ is\ symmetric\, .}
\end{array}
\right.
\ee
where we assumed that $\kappa$ is large enough so that $\frac{\mathrm d^2}{\mathrm d z^2}\hat e_{T,\kappa}^+ (z)=0$.
For the symbol of the XY Hamiltonian this condition is already satisfied for $\kappa= 1$.  

Some properties of the solutions of \eref{eq:sys} will be analyzed in \sref{s:TH}.
In principle the identification of all the local conservation laws requires the solution of the problem for generic $\kappa$. However, in the simplest cases only the smallest values of $\kappa$ give independent solutions.
Our analysis evinces the following:
\begin{itemize}
\item[-] In the XY model ($h=0$), setting $\kappa=2$ is sufficient to obtain all the solutions;
\item[-] In the transverse-field Ising chain ($\gamma=1$), one can choose $\kappa=1$;
\item[-] In the XY model in a transverse field ($h\neq0$ and $\gamma\neq 1$), we do not find an infinite number of solutions at fixed $\kappa$ (at least analyzing the smallest values of $\kappa$). This manifests a weakness in the Hankel structure that we are imposing, which is not effective in capturing the symmetries of the boundary part of the charges. Nevertheless, in \sref{s:XY} we propose a conjecture for the symbol of the Toeplitz part of the charges that are invariant under chain inversion (\emph{cf}. \eref{eq:conj0}). 
\end{itemize}
Remarkably, using reflection symmetry, in \sref{ss:refsym} we show that, at fixed $\kappa$, almost any solution of  \eref{eq:sys} is also solution of the following simplified system of equations:
\be\label{eq:redsys001}
\left\{
\begin{array}{l}
[\hat e_{T,\kappa}(z),\hat q_{T,\kappa}(z)]=0\\
\Bigl(\frac{1}{z}\hat q_{T,\kappa}^+(z)-\hat q_{H,\kappa}(z)\Bigr)\hat e_{T,\kappa}^-=0\\
\Bigl[\hat e_{T,\kappa}(z),\hat q_{H,\kappa}(z)\Sigma_\kappa^x\otimes \sigma^y\Bigr]= 0\, ,
\end{array}
\right.
\ee
where $\Sigma_\kappa^x$ is the $\kappa$-by-$\kappa$ exchange matrix $[\Sigma_\kappa^x]_{i j}=\delta_{i+j,\kappa+1}$. 
\subsection{Quasilocalized conserved quantities}\label{eq:ssq-loc}
In the ferromagnetic phase of the XY model in a transverse field ($|h|<1$ and $\gamma\neq 0$) there is a conserved Majorana fermion $B^L$ quasilocalized at the boundary~\cite{Kitaev}. 
With zero transverse field this is given by
\be\label{eq:BLXY}
B^L=\sqrt{2\sinh(2\alpha)}\sum_{n=1}^{\infty} (-1)^n e^{-(2n-1)\alpha}\prod_{j<2n-1}\sigma_j^z\ \sigma_{2n-1}^x\qquad (h=0)\, .
\ee
In the transverse-field Ising model it reads as
\be\label{eq:BLTFIC}
B^L=\sqrt{1-h^2}\sum_{n=1}^\infty h^{n-1}\prod_{j<n}\sigma_j^z\ \sigma_n^x\qquad (\gamma=1)\, .
\ee
More generally, in the quantum XY model in a transverse field it is given by
\be\label{eq:BLXYh}
B^L=\frac{\sqrt{\gamma(1-h^2)}}{\zeta}\sum_{n=1}^\infty \Bigl[\Bigl(\frac{1-\gamma}{h+\zeta}\Bigr)^{n}-\Bigl(\frac{1-\gamma}{h-\zeta}\Bigr)^{n}\Bigr]\prod_{j<n}\sigma_j^z\ \sigma_n^x\, ,
\ee
where $\zeta=\sqrt{h^2+\gamma^2-1}$. 
The privileged direction $\hat x$ comes from having considered $\gamma$ positive (in particular, the ground state of the model is ferromagnetic along $x$). 
Apart from being quasilocalized at the boundary, this conservation law has the peculiarity of anticommuting with $\Pi^z$. 
\paragraph{General formalism.}
When there is a boundary bound state, the resulting quasilocalized conserved quantity corresponds to an eigenvector with zero eigenvalue of the block-Toeplitz operator $\mathcal H$ \eref{eq:blockT}.  

We use a minimal ansatz and seek for a  solution of the form 
\begin{equation*}
[\vec{\mathcal V}]_{2\kappa \ell+j}= [e^{-W \ell}\vec v]_j\qquad j=1,\dots, 2\kappa\, ,\qquad \ell=0,1,\dots
\end{equation*}
where quasilocality implies that the eigenvalues of $W$ have strictly positive real part. If we apply $\mathcal H$ to this vector and impose that the eigenvalue is zero, after integration in the complex plane we obtain
\bea\label{eq:boundsys0}
(\hat e_{T,\kappa}^-\hat e^{-W}+\hat e_{T,\kappa}^0)\vec v=0\nn
(\hat e_{T,\kappa}^- e^{-W}+\hat e_{T,\kappa}^0 +\hat e_{T,\kappa}^+ e^{W})e^{-\ell W}\vec v=0\qquad\forall (\mathbb N\ni)\ell\geq 1
\eea
In particular, this system has a solution if $\vec v$ is an eigenstate of $e^{-W}$ and is in the kernel of $\hat e_{T,\kappa} (e^{w})$, where $e^{W}\vec v=e^w\vec v$ ($\mathrm{Re}[w]>0$). We then obtain
\bea\label{eq:boundsys}
\hat e_{T,\kappa}^+\vec v_w=0\nn
\hat e_{T,\kappa} (e^{w})\vec v_w=0\, .
\eea
Both for $\gamma=0$ \eref{eq:BLXY} and in the transverse-field Ising chain ($\gamma=1$) \eref{eq:BLTFIC}, the quasilocalized conserved quantity associated with the bound state is the solution of \eref{eq:boundsys} for $\kappa=2$ and $\kappa=1$, respectively. The XY model in a transverse field is slightly more complicated. Indeed, for $\kappa=2$, the first equation of \eref{eq:boundsys} is not solved by the vector in the kernel of $\hat e_{T,\kappa} (e^{w})$ (for the particular values of $w$ for which the equation has a solution). In similar situations one must consider a generic linear combination of the vectors $\vec v_{w_i}$ satisfying the second equation of \eref{eq:boundsys}
\begin{equation*}
\vec v=\sum_{i=1} c_i \vec v_{w_i}\, .
\end{equation*} 
Imposing the first equation of \eref{eq:boundsys0}
\begin{equation*}
\hat e_{T,\kappa}^+\sum_{i=1} c_ie^{w_i}\vec v_{w_i}=0
\end{equation*}
one finally obtains
\begin{equation*}
[\vec{\mathcal V}]_{2\kappa \ell+j}= \sum_{i=1}c_i e^{-w_i \ell}[\vec v_{w_i}]_j\, .
\end{equation*}
Inspecting \eref{eq:BLXYh}, the reader can realize that the quasilocalized conserved quantity in the XY model in a transverse field displays this slightly more complicated structure.  

\subsection{Remarks}
We would like to emphasize some strengths and weaknesses of \eref{eq:sys} (especially in the simplified version \eref{eq:redsys001}) and \eref{eq:boundsys0}. We started with a Hamiltonian of a quantum many-body system. The mapping to noninteracting fermions allowed to reduce the complexity of the problem exponentially but, in fact, in the thermodynamic limit we still had to work with operators. Nevertheless, the systems of equations that we presented are written in terms of finite matrix functions and can be solved systematically, varying the block's size $\kappa$ if necessary. In addition, there is a standard way (see \sref{s:TH}) to generate infinitely many solutions associated with conserved quantities with increasing range from a given solution of the simplified system of equations \eref{eq:redsys001}.

Although the equations have been derived assuming locality, one can relax locality into quasilocality by replacing the assumption that the symbols are polynomials of $z=e^{ik}$ and $1/z$ with analyticity in the unit circle.  

A weakness of \eref{eq:sys} is in the implicit assumption that the boundary term has  Hankel structure. If this is not the case, the block's size $\kappa$ associated with the conservation law will be at least equal to the range of the boundary term. This means that one is forced to solve the system for arbitrary $\kappa$, reducing the effectiveness of the approach (see \sref{s:XY}).

The worst scenario is the presence of conservation laws that are quasilocal in the bulk and have unstructured boundary terms: without further manipulations, these solutions are likely to be missed.
 
\begin{figure}
\begin{center}
\includegraphics[width=0.7\textwidth]{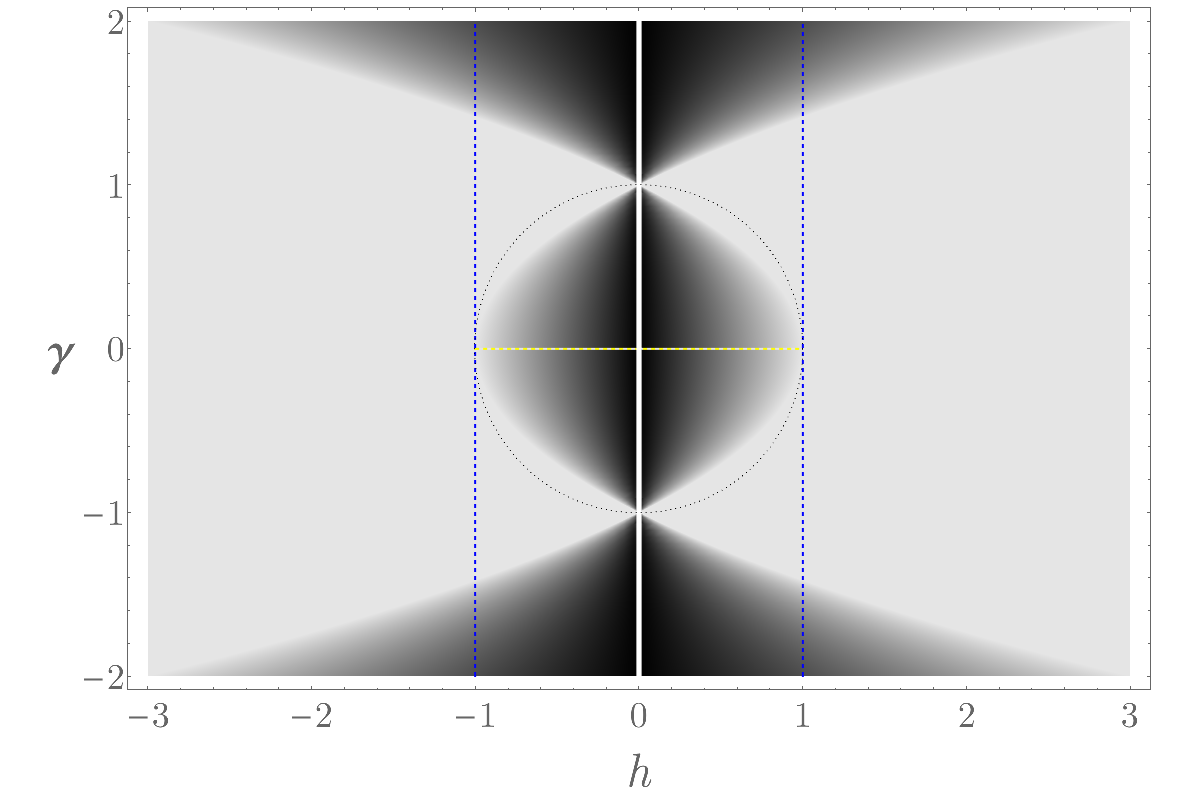}
\caption{Cartoon of the quasilocal conservation laws that we identified in the quantum XY model in a transverse field. The light gray region corresponds to the ``standard'' situation where any reflection symmetric charge of the infinite chain survives a (free) boundary. The region becomes darker  as the number of charges is reduced and black means that we found only half of the expected number of conservation laws. The white thick line for $h=0$ indicates the presence of non-commuting charges and that, in turn, the set is practically two times as large as in the ``standard'' case. The dashed lines indicate the (conformal) critical regions (yellow means central charge $c=1$ and blue $c=\frac{1}{2}$) and the dotted circle is another relevant curve of the phase diagram.}\label{f:phasediag}
\end{center}
\end{figure}

\Fref{f:phasediag} depicts our current understanding of the quasilocal charges of the quantum XY model in a transverse field with OBC as a function of the Hamiltonian parameters. We expect that the structure of the charges depends only on the ratio $q=\frac{h}{\gamma^2-1}$ and, for $|q|<1$, 
the smaller $|q|$ is and the smaller the set of quasilocal conserved quantities. This rule has the exceptions $q=0$, where the set of local charges doubles because the model becomes non-abelian integrable~\cite{F:super}, and $\gamma=0$, where the situation is analogous to $|q|>1$.

\section{Excitations for $h=0$}\label{s:excXY}
Following \cite{LSM:XY}, the Hamiltonian of the XY model with open boundary conditions was diagonalized in \cite{KarevskiBook}. For $h=0$ we detail the solution of the model in \ref{a:XY0}. The reader interested also in the diagonalization of the transverse-field Ising chain and of the quantum XY model in a transverse field can find some details in \ref{a:TFIC} and \ref{a:XYh}, respectively. Incidentally, we are not aware of any work where the quantum XY model in a transverse field with OBC is diagonalized properly. 

By means of \eref{eq:bU}, the excitations of the model  follow directly from the eigenvectors of $\mathcal H$ \eref{eq:blockT},  which for $h=0$ are computed in \ref{a:XY0} and given by  \eref{eq:evec}. 
We have identified two families of creation operators. 
We use a subscript $\pm$ to refer to the one or to the other. 

\subsection{Even chain}
The excitations are given by (\emph{cf.} \eref{eq:a})
\begin{eqnarray}\label{eq:ext1}
\fl\qquad&b_+^\dag(\phi)=\sum_{n=1}^{L/2} \frac{\left[\sin(n\phi)e^{-\alpha} +\sin((n-1) \phi)e^{\alpha}\right]a_{2n-1}^x+i \varepsilon(\varphi)\sin(n\phi) a_{2n}^y}{\sqrt{(L+1)\varepsilon^2(\varphi)- 2\sinh(2\alpha)}}\\\label{eq:ext2}
\fl\qquad &b^\dag_-(\phi)=\sum_{n=1}^{L/2} \frac{\left[\sin(n\phi)e^{\alpha} +\sin((n-1) \phi)e^{-\alpha}\right]a_{2n-1}^y-i \varepsilon(\varphi)\sin(n\phi_-)a_{2n}^x}{\sqrt{(L+1)\varepsilon^2(\varphi)+ 2\sinh(2\alpha)}}\, ,
\end{eqnarray}
where $\alpha=\mathrm{arctanh}\gamma$ \eref{eq:paramXY} and the pseudomomenta $\phi$ satisfy the quantization conditions
\be\label{eq:quantEven}
e^{i(L+1)\phi}=\frac{\cosh(\alpha\pm i\frac{\phi}{2})}{\cosh(\alpha\mp i\frac{\phi}{2})}\, .
\ee
The corresponding excitation energies are
\be\label{eq:epsilon}
\varepsilon(\phi)=2\sqrt{\cos^2\frac{\phi}{2}+\sinh^2\alpha}\, .
\ee 
Since the pseudomomenta associated with the two types of excitations are different, their energy is non-degenerate.
From the excitations we can construct the conserved occupation numbers $n_s(\phi)=b^\dag_s(\phi) b_s(\phi)$.

\Tref{t:sym} shows the transformation rules for the excitations and for the occupation numbers under the symmetry transformations of the Hamiltonian. 
\begin{table}
\begin{center}
\begin{tabular}{l|c|c|c|c}
&$\bf R$&$\Pi^x$&$\Pi^y$&$\Pi^z$\\
\hline
\textsc{even chain}&\\
\hline
$b^\dag_s(\phi)$&$-(-1)^{[\phi]}\Pi^z b^\dag_s(\phi) $&$s b^\dag_s(\phi)$&$-s b^\dag_s(\phi)$&$- b^\dag_s(\phi)$\\
$n_s(\phi)\equiv b^\dag_s(\phi) b^{\phantom \dag}_s(\phi)$&$n_s(\phi)$&$n_s(\phi)$&$n_s(\phi)$&$n_s(\phi)$\\
\hline
\textsc{odd chain}&\\
\hline
$\tilde b^\dag_s(\phi)$&$-(-1)^{[\phi]}\Pi^z \tilde b^\dag_{-s}(\phi) $&$s \tilde b^\dag_s(\phi)$&$-s\tilde b^\dag_s(\phi)$&$-\tilde b^\dag_s(\phi)$\\
$\tilde n_s(\phi)\equiv \tilde b^\dag_s(\phi) \tilde b^{\phantom \dag}_s(\phi)$&$\tilde n_{-s}(\phi)$&$\tilde n_s(\phi)$&$\tilde n_s(\phi)$&$\tilde n_s(\phi)$\\
$w_s(\phi)\equiv e^{i\frac{1-s}{2}\frac{\pi}{2}}\tilde b^\dag_{+}(\phi)\tilde b^{\phantom \dag}_{-}(\phi)+$h.c.&$s \tilde w_s(\phi)$&$-w_s(\phi)$&$-w_s(\phi)$&$w_s(\phi)$\\
\end{tabular}
\caption{Transformation of the excitations and of the elementary conserved quantities of the XY model ($h=0$) under chain inversion $\bf R$ and spin flip symmetry $\Pi^\alpha$ about axis $\alpha\in\{x,y,z\}$. Index $s=\pm 1$ identifies the family of excitations. Integer $[\phi]$ is defined just below \eref{eq:transf}.}\label{t:sym}
\end{center}
\end{table}

\subsubsection*{Bound state.}
As shown in \ref{a:XY0}, the first family of excitations includes  a bound state with a complex pseudomomentum \eref{eq:BS} ($\phi=\pi+2i\alpha-O(e^{-2L\alpha})$). 
Its energy vanishes exponentially in the thermodynamic limit and is approximately given by
\begin{equation*}
\varepsilon_B
\sim 2\sinh(2\alpha) e^{-(L+1)\alpha}\, .
\end{equation*}

Let us define the Majorana fermions
\begin{eqnarray}\label{eq:BSMajo}
&B^L=b_+^\dag(\pi+2i\alpha^-)+b_+(\pi+2i\alpha^-)\\
&B^R=i[b_+(\pi+2i\alpha^-)-b_+^\dag(\pi+2i\alpha^-)]\, ,
\end{eqnarray}
which satisfy the algebra
\begin{equation*}
i[H,B^{L,R}]=\varepsilon_B B^{R,L}\, .
\end{equation*} 
It turns out that $B^L$ is quasilocalized around the left boundary and, for large $L$, reads
\be\label{eq:BL}
B^L\sim \sqrt{2\sinh(2\alpha)}\sum_{n=1}^{L/2} (-1)^n e^{-(2n-1)\alpha}a_{2n-1}^x
\ee
which in the spin representation becomes
\begin{equation*}
B^L\sim\sqrt{2\sinh(2\alpha)}\sum_{n=1}^{L/2} (-1)^n e^{-(2n-1)\alpha}\prod_{j<2n-1}\sigma_j^z\ \sigma_{2n-1}^x\, .
\end{equation*}
Analogously, we have
\be\label{eq:BR}
B^R\sim\sqrt{2\sinh(2\alpha)}\sum_{n=1}^{L/2} (-1)^n e^{-(L-2n+1)\alpha}a_{2n}^y\, .
\ee
There is a subtlety related to the nonlocality of the Jordan-Wigner transformation. The Majorana fermion $B^R$ is nonlocal in the spin representation:
\begin{equation*}
B^R\sim \sqrt{2\sinh(2\alpha)}\sum_{n=1}^{L/2} (-1)^n e^{-(L-2n+1)\alpha}\prod_{j<2n}\sigma_j^z\ \sigma_{2n}^y\, .
\end{equation*}
In fact, this kind of nonlocality can be easily cured by applying the spin-flip operator $\Pi^z$ \eref{eq:Piz}. We can indeed define the Majorana fermion
\begin{equation*}
\bar B^R=i \Pi^z B^R\sim\sqrt{2\sinh(2\alpha)}\sum_{n=1}^{L/2} (-1)^n e^{-(L-2n+1)\alpha} \sigma_{2n}^x \prod_{j> 2n}\sigma_j^z
\end{equation*}
which is instead quasilocalized at the right boundary. We point out that $\bar B^R$ \emph{commutes} with $B^L$, as one would expect for operators that act nontrivially far away from each other. In fact, it also commutes with all the other excitations.

\subsection{Odd chain}\label{s:BSodd}
If $L$ is odd, there are still two families of excitations but, in both cases, the pseudomomenta have the same quantization rules.  
In a particular basis, the excitations read
\begin{eqnarray}\label{eq:tildebOdd}
\fl\qquad&\tilde b^\dag_+(\phi)=\sum_{n=1}^{(L+1)/2} \frac{\left[\sin(n\phi)e^{-\alpha} +\sin((n-1) \phi)e^{\alpha}\right]a_{2n-1}^x}{\sqrt{L+1}\varepsilon(\phi)}+i\sum_{n=1}^{(L-1)/2}\frac{\sin(n\phi) a_{2n}^y}{\sqrt{L+1}}\nonumber\\
\fl\qquad &\tilde b^\dag_-(\phi)=\sum_{n=1}^{(L+1)/2} i \frac{\left[\sin(n\phi)e^{\alpha} +\sin((n-1) \phi)e^{-\alpha}\right]a_{2n-1}^y}{\sqrt{L+1}\varepsilon(\phi)}+\sum_{n=1}^{(L-1)/2}\frac{\sin(n\phi) a_{2n}^x}{\sqrt{L+1}}\, ,
\end{eqnarray}
where the pseudomomenta $\phi$ are quantized according to
\begin{equation}\label{eq:QCodd1}
\phi=\frac{2\pi n}{L+1}\qquad n=1,\dots,\frac{L-1}{2}\, ,
\end{equation}
for the real solutions, and
\begin{equation*}
\phi=\pi+2i\alpha\, ,
\end{equation*}
for the bound state. 

The excitation energy $\varepsilon(\phi)$  has the same functional form as in the even case \eref{eq:epsilon}. Having the two families of excitations identical quantization conditions,  $b^\dag_+(\phi)$ and $b^\dag_-(\phi)$ have the same energy and can be mixed
\begin{equation*}
\left(\begin{array}{c}
\tilde b^\dag_+(\phi)\\
\tilde b^\dag_-(\phi)
\end{array}\right)\rightarrow U\left(\begin{array}{c}
\tilde b^\dag_+(\phi)\\
\tilde b^\dag_-(\phi)
\end{array}\right)\, ,
\end{equation*}
where $U$ is an arbitrary 2-by-2 unitary matrix.
Thus, the occupation numbers in a generic basis are linear combinations of 
\begin{eqnarray}\label{eq:ONodd}
\tilde n_\pm (\phi)=\tilde b_\pm^\dag(\phi)\tilde b_\pm^{\phantom\dag}(\phi)\nn
w_+(\phi)=\tilde b^\dag_{+}(\phi)\tilde b_{-}(\phi)+\tilde b^\dag_{-}(\phi)\tilde b_{+}(\phi)\nn
w_-(\phi)=i\bigl(\tilde b^\dag_{+}(\phi)\tilde b_{-}(\phi)-\tilde b^\dag_{-}(\phi)\tilde b_{+}(\phi)\bigr)\, .
\end{eqnarray}
In \tref{t:sym} we report the transformation rules for the excitations and for the elementary conserved quantities.

\subsubsection*{Bound state.}

The bound state has exactly  zero energy. 
We find the \emph{exactly conserved} Majorana fermions
\begin{eqnarray}\label{eq:BLRodd}
B^L= \sqrt{\frac{\sinh(2\alpha)}{\sinh(\alpha(L+1))}}e^{\alpha\frac{L+1}{2}}\sum_{n=1}^{(L+1)/2} (-1)^n e^{-(2n-1)\alpha}a_{2n-1}^x\nn
B^R=\sqrt{\frac{\sinh(2\alpha)}{\sinh(\alpha(L+1))}}e^{\alpha\frac{L+1}{2}}\sum_{n=1}^{(L+1)/2} (-1)^n e^{-(L+2-2n)\alpha}a_{2n-1}^y\, .
\end{eqnarray}
Again, $B^L$ and $\bar B^R=i\Pi^z B^R$ are quasilocalized around the left and right boundaries, respectively.

\section{Local charges for $h=0$: direct approach}\label{s:lclXY}\label{s:finf}

Not any local or quasilocal conservation law has a finite volume analogue. 
This is related to the fact that finite and infinite chains can have different symmetries and in the thermodynamic limit additional \mbox{(quasi)local} conservation laws can emerge.
In this section we will consider finite chains and identify the charges that remain local in the thermodynamic limit. 
With the aim of being didactic, we will follow a direct approach that allows one to easily identify the various families of local conserved quantities. In the simplest (and more standard) cases we will construct the charges explicitly. Otherwise, we will rely on the systematic procedure described in the following. 

As pointed out in Ref.~\cite{GM:open}, a local conservation law $Q$ can be written as the sum of a bulk term, which has the same density of a local charge in the infinite chain, and two boundary terms, which are localized around the left and the right boundaries. 
Incidentally, the bulk part can also vanish; in that case the conserved quantity is \mbox{(quasi)localized} at the boundaries. 
Assuming that the boundary parts have a finite range, the conservation laws of the finite chain can be determined as follows:
\begin{enumerate}
\item The structure of the candidate local conservation law is identified in the finite chain. 
\item Since the Hamiltonian is invariant under chain inversion, the chain-reversed conservation law is either itself (up to the sign) or another local charge. Such transformation rules are determined in the finite chain. 
\item 
Being the conservation law local, the left boundary term and the bulk part of the conservation law can be worked out in the thermodynamic limit. 
\item The right boundary term is the left boundary term of the chain-reversed conservation law, which corresponds to the same or to an other local charge of the semi-infinite chain.
\item The conservation law in the finite chain can then be reconstructed.
\end{enumerate}
We would like to point out two reasons why it is worth considering  finite systems:
\begin{itemize}
\item[-] In the thermodynamic limit it is much more difficult to gain a perception of the ``completeness'' of the set of conservation laws that one tries and construct.
\item[-] 
The stationary properties of observables after global quenches are sometimes extracted from diagonal ensembles, which correspond to infinite time averages in finite chains in the limit of infinite chain's length $L$. 
If there are families of \mbox{(quasi)local} conservation laws that depend nontrivially on  $L$, the diagonal ensemble is ill-defined and such descriptions can not be used, unless the structure of the conservation laws for finite $L$ is clearly known. 
\end{itemize}

\subsection{Even chain}\label{s:clEven}
When $L$ is even, the different quantization rules for the two families of excitations 
\eref{eq:ext1} and \eref{eq:ext2} result in non-degenerate excitation energies. As a consequence, among the noninteracting operators, only linear combinations of the occupation numbers are conserved. The occupations numbers
\begin{equation*}
n_\pm(\phi)-\frac{1}{2}=\frac{1}{2}(b^\dag_\pm(\phi)-b_\pm(\phi))(b^\dag_\pm(\phi)+b_\pm(\phi))\, ,
\end{equation*}
here shifted by a constant, are generally nonlocal operators.  An exception is given by
\be
C_b=\Pi^z(2n_{+}(\pi+2i\alpha^-)-1)=i \bar B^R B^L\, ,
\ee
which is quasilocalized around the boundaries and can be interpreted as a quasilocal conservation law for the spin Hamiltonian with the two boundaries wrapped so as to be close to each other.

We now work out the \emph{local} conservation laws (with range independent of $L$). 
The general procedure is to consider a sum of the form
\begin{equation*}
Q_f^{\pm}=\sum_\phi f(\phi)(n_\pm(\phi)-1/2)
\end{equation*}
and choose $f$ so that the orthogonality relations \eref{eq:ortE1}\eref{eq:ortE2} bound the range of the operator from above. 
More explicitly we have
\begin{eqnarray}\label{eq:QfEven}
\fl \qquad Q_f=\frac{1}{2}\sum_\phi f(\phi)(b^\dag_\pm(\phi)-b_\pm(\phi))(b^\dag_\pm(\phi)+b_\pm(\phi))\nonumber\\
\fl\qquad\qquad=\pm 2i\sum_\phi \frac{f(\phi)}{\varepsilon(\varphi)}\sum_{\ell, n=1}^{L/2} \frac{\sin(n\phi) \left[\sin(\ell\phi)e^{\mp\alpha} +\sin((\ell-1) \phi)e^{\pm\alpha}\right]a_{2n}^{y(x)} a_{2\ell-1}^{x(y)}}{L+1\mp \frac{2\sinh(2\alpha)}{\varepsilon^2(\varphi)}}\, .
\end{eqnarray}
We can already notice that the range of the operator is always even, that is to say, no (quadratic) operator with odd range can be exactly conserved. 

Since the occupation numbers are invariant under chain inversion,  we can 
rewrite the operator as
\be
Q_f^{\pm}=C_f^{\pm}+{\bf R}C_f^{\pm} {\bf R}+M_f^{\pm}
\ee
with
\begin{equation*}
\fl C_f^{\pm}=\pm 2i\sum_\phi \frac{f(\phi)}{\varepsilon(\varphi)}\sum_{\ell, n=1\atop \ell+n\leq \frac{L}{2}}^{L/2-1}\frac{\sin(n\phi) \left[\sin(\ell\phi)e^{\mp\alpha} +\sin((\ell-1) \phi)e^{\pm\alpha}\right]a_{2n}^{y(x)} a_{2\ell-1}^{x(y)}}{L+1\mp \frac{2\sinh(2\alpha)}{\varepsilon^2(\varphi)}}\, ,
\end{equation*}
\begin{equation*}
\fl M_f^{\pm}=\pm 2i \sum_{\phi, \ell} \frac{f(\phi)}{\varepsilon(\varphi)} \frac{\sin\bigl((\frac{L}{2}+1-\ell)\phi\bigr) \left[\sin(\ell\phi)e^{\mp\alpha} +\sin((\ell-1) \phi)e^{\pm\alpha}\right]a_{L+1-2\ell}^{y(x)} a_{2\ell-1}^{x(y)}}{L+1\mp \frac{2\sinh(2\alpha)}{\varepsilon^2(\varphi)}}\, .
\end{equation*}
We report here for convenience the orthogonality relation \eref{eq:ortE1}:
\be\label{eq:ortE1rep}
\sum_{\phi_{(s)}}\frac{\sin(\ell \phi)\sin(n \phi)}{L+1-2s\frac{\sinh(2\alpha)}{\varepsilon^2(\phi)}}=\frac{1}{4}\delta_{\ell n}\, .
\ee
This is the basic ingredient to determine the local conservation laws. In particular, comparing \eref{eq:ortE1rep} with $C_j^{\pm}$ and $M_f^{\pm}$, we easily deduce that the operators are local when $f_j(\phi)=\varepsilon(\phi)\cos(j\phi)$. This can be seen as follows. Let us focus on $C_{f_j}$ 
\begin{equation*}
\fl \ C_{f_j}^{\pm}=\pm2i\sum_\phi\sum_{\ell,n=1\atop\ell+n\leq \frac{L}{2}}^{L/2-1} \frac{\cos(j \phi)\sin(n\phi) \left[\sin(\ell\phi)e^{\mp\alpha} +\sin((\ell-1) \phi)e^{\pm\alpha}\right]a_{2n}^{y(x)} a_{2\ell-1}^{x(y)}}{L+1\mp \frac{2\sinh(2\alpha)}{\varepsilon^2(\varphi)}}\, .
\end{equation*}
If $\ell\leq n$, we can use the Prosthaphaeresis formula for $\cos(j \phi)\sin(\ell\phi)$ and $\cos(j \phi)\sin((\ell-1)\phi)$, whereas for $n< \ell $ we can use the transformation for $\cos(j \phi)\sin(n\phi)$. In this way, if $j\leq \frac{L}{4}$, the arguments of the sin's are always of the form $m\phi$, with $m$ smaller than or equal to $\frac{L}{2}$, as required in \eref{eq:ortE1rep}. 
Thus, the difference between $\ell$ and $n$ is constrained to be smaller or equal to $j$, which in turn determines the range of the operator. 
The same argument applies to $M^{\pm}_{f_j}$, proving that $Q_{f_j}$ is local.
Let us work out it explicitly. 
At fixed $j$, we consider $L$ large enough that there is no contribution from the terms that are nonzero only if $j$ is comparable with $L$. We obtain
\be\label{eq:Ideven}
\fl \quad C^{\pm}_{f_j}=\pm \frac{i}{4}\sum_{\ell,n=1\atop\ell+n\leq \frac{L}{2}}^{L/2-1} (\delta_{\ell-n, j}+\delta_{\ell-n, -j}-\delta_{\ell+n, j})(e^{\mp\alpha} a_{2n}^{y(x)} a_{2\ell-1}^{x(y)}+e^{\pm \alpha} a_{2n}^{y(x)} a_{2\ell+1}^{x(y)})
\ee
The contribution from $M^{\pm}_{f_j}$ consists of a few operators localized around half chain with range $\sim 2j$ and has the same form of the terms that, in the previous expression, depend on the difference of indices.
This results in 
\begin{eqnarray}\label{eq:Q}
\fl \quad Q^{\pm}_{f_j}=\pm \frac{i e^{\mp \alpha}}{4}\sum_{\ell=1}^{\frac{L}{2}-j}a_{2\ell}^{y(x)} a_{2\ell+2j -1}^{x(y)}-a_{2\ell-1}^{x(y)}a_{2\ell+2j }^{y(x)}\pm \frac{i e^{\pm \alpha}}{4}\sum_{\ell=1}^{\frac{L}{2}-j-1} a_{2\ell}^{y(x)} a_{2\ell+2j+1}^{x(y)} -a_{2\ell+1}^{x(y)} a_{2\ell+2j}^{y(x)} \nonumber\\
\mp\frac{i}{4}\sum_{\ell=1}^{j-1} (e^{\mp\alpha} a_{2j-2\ell}^{y(x)} a_{2\ell-1}^{x(y)}+e^{\pm \alpha} a_{2j-2\ell}^{y(x)} a_{2\ell+1}^{x(y)})\nonumber\\
\qquad \mp\frac{i}{4}\sum_{\ell=1}^{j-1} (e^{\mp\alpha} a_{L+2-2\ell}^{y(x)} a_{L+1-2j+2\ell}^{x(y)} +e^{\pm \alpha} a_{L-2\ell}^{y(x)} a_{L+1-2j+2\ell}^{x(y)} )\, .
\end{eqnarray}
In particular, for $j=0$ we find
\begin{equation*}
\quad Q^{\pm}_{f_0}=\mp \frac{i}{2}\sum_{\ell=1}^{L/2}e^{\mp\alpha}  a_{2\ell-1}^{x(y)}a_{2\ell}^{y(x)}\pm \frac{i}{2}\sum_{\ell=1}^{L/2-1}e^{\pm\alpha}a_{2\ell}^{y(x)} a_{2\ell+1}^{x(y)}\, .
\end{equation*}
As expected, the Hamiltonian is the sum of the two charges for $j=0$
\begin{equation*}
H=Q_{f_0}^++Q_{f_0}^-\, .
\end{equation*} 
More generally, the local conservation laws that are one-site shift invariant in the bulk are given by
\begin{equation}
I_j^{+(e)}\bigl(=Q_{f_j}^++Q_{f_j}^-\bigr)= I_j^{+(e);\, bulk}+I_j^{+(e);\, edges}
\end{equation}
where
\begin{equation*}
\fl\qquad I_j^{+(e);\, bulk}=
\frac{i}{4}\sum_{\ell=2}^{L-2j}e^{-\alpha}(a_{\ell}^{y} a_{\ell+2j -1}^{x}-a_{\ell-1}^{x} a_{\ell+2j}^{y})+e^{\alpha}(a_{\ell-1}^y a_{\ell+2j}^x-a_{\ell}^xa^y_{\ell+2j-1})
\end{equation*}
\begin{eqnarray*}
\fl\qquad I_j^{+(e);\, edges}=\frac{i}{4}\sum_{\ell=1}^{j-1} \Bigl[e^{- \alpha} (a_{2j-2\ell}^{x} a_{2\ell+1}^{y}+a_{2\ell-1}^{x}a_{2j-2\ell}^{y} )+e^{\alpha} (a_{2j-2\ell}^{x} a_{2\ell-1}^{y}+ a_{2\ell+1}^{x}a_{2j-2\ell}^{y})\nonumber\\
+ e^{\alpha} (a_{L+2-2\ell}^{x} a_{L+1-2j+2\ell}^{y}-a_{L-2\ell}^{y} a_{L+1-2j+2\ell}^{x}) \nonumber\\
\qquad\qquad+e^{- \alpha} (a_{L-2\ell}^{x} a_{L+1-2j+2\ell}^{y}-a_{L+2-2\ell}^{y} a_{L+1-2j+2\ell}^{x})\Bigr]\, .
\end{eqnarray*}
The remaining independent conservation laws are given by
\be\label{eq:JEven}
J_j^{+(e)}\bigl(=Q_{f_j}^+-Q_{f_j}^-\bigr)= J_j^{+(e);\, bulk}+J_j^{+(e);\, edges}
\ee
where
\be
\fl\quad J_j^{+(e);\, bulk}=\frac{i}{4}\sum_{\ell=2}^{L-2j}(-1)^\ell[e^{-\alpha}(a_{\ell}^{y} a_{\ell+2j -1}^{x}-a_{\ell-1}^{x} a_{\ell+2j}^{y})-e^{\alpha}(a_{\ell-1}^y a_{\ell+2j}^x-a_{\ell}^xa^y_{\ell+2j-1})]
\ee
\begin{eqnarray*}
\fl\quad J_j^{+(e);\, edges}=-\frac{i}{4}\sum_{\ell=1}^{j-1} \Bigl[e^{- \alpha} (a_{2j-2\ell}^{x} a_{2\ell+1}^{y}+a_{2j-2\ell}^{y}a_{2\ell-1}^{x} )+e^{\alpha} (a_{2j-2\ell}^{x} a_{2\ell-1}^{y}+a_{2j-2\ell}^{y}a_{2\ell+1}^{x})\nonumber\\
+e^{- \alpha} (a_{L-2\ell}^{x} a_{L+1-2j+2\ell}^{y}+a_{L+2-2\ell}^{y} a_{L+1-2j+2\ell}^{x})\nonumber\\
\qquad\qquad +e^{\alpha} (a_{L+2-2\ell}^{x} a_{L+1-2j+2\ell}^{y}+a_{L-2\ell}^{y} a_{L+1-2j+2\ell}^{x})\Bigr]\, .
\end{eqnarray*}

The explicit expressions in terms of spin variables can be readily obtained from the Jordan-Wigner transformation \eref{eq:a}. In particular we have ($n>0$)
\begin{eqnarray*}
i a_\ell^x a_{\ell+n}^y&=\sigma_\ell^y\sigma_{\ell+1}^z\cdots \sigma_{\ell+n-1}^z\sigma_{\ell+n}^y\\
i a_\ell^y a_{\ell+n}^x&=-\sigma_\ell^x\sigma_{\ell+1}^z\cdots \sigma_{\ell+n-1}^z\sigma_{\ell+n}^x\, .
\end{eqnarray*}

\subsection{Odd chain}
First of all we remind the reader that, as observed in \sref{s:BSodd}, the Majorana fermions $B^L$ and $\bar B^R$ \eref{eq:BLRodd} are conserved in the finite chain. 
These are quasilocal conservation laws localized at the left and right boundaries.

We can then focus on the conservation laws that are written in terms of real pseudomomenta.
Among the noninteracting operators, the four independent classes of exactly conserved quantities are reported in \eref{eq:ONodd}. 
Following the same method used in the even case, the first two families of local conserved quantities are given by:
\begin{eqnarray*}
\tilde Q_{f_j}^{\pm}=\frac{1}{2}\sum_\phi \varepsilon(\phi)\cos(j\phi)(\tilde b^\dag_\pm(\phi)-\tilde b_\pm(\phi))(\tilde b^\dag_\pm(\phi)+\tilde b_\pm(\phi))\nonumber\\
\fl\quad=\pm 2i\sum_{k=1}^{\frac{L-1}{2}} \sum_{\ell=1}^{\frac{L+1}{2}} \sum_{n=1}^{\frac{L-1}{2}} \cos(j\phi_k)\frac{\sin(n\phi_k) \left[\sin(\ell\phi_k)e^{\mp\alpha} +\sin((\ell-1) \phi_k)e^{\pm\alpha}\right]a_{2n}^{y(x)} a_{2\ell-1}^{x(y)}}{L+1}
\end{eqnarray*}
where $\phi_k=\frac{2\pi k}{L+1}$. Due to the simple quantization conditions, in this case we can perform the sums directly and use the following identity
\be\label{eq:indentity}
\fl\quad \frac{1}{L+1}\sum_k \cos(j\phi_k)\sin(n\phi_k) \sin(\ell \phi_k)=\frac{\delta_{\ell-n, j}+\delta_{\ell-n, -j}-\delta_{\ell+n, j}-\delta_{\ell+n, L+1-j}}{8}\, ,
\ee
valid for $j$ sufficiently smaller than $L$.
We note that this is essentially the same relation found in the even case (\emph{cf.} \eref{eq:Ideven}). 
We find
\begin{eqnarray}\label{eq:tildeQodd}
\fl\ \tilde Q_{f_j}^{\pm}=
\pm \frac{i}{4}\sum_{\ell=1}^{\frac{L-1}{2}} e^{\mp\alpha}(a_{2\ell}^{y(x)}a_{2\ell+2j-1}^{x(y)}-a_{2\ell-1}^{x(y)}a_{2\ell+2j}^{y(x)})+e^{\pm\alpha}(a_{2\ell}^{y(x)}a_{2\ell+2j+1}^{x(y)}-a_{2\ell+1}^{x(y)}a_{2\ell+2j}^{y(x)})\nonumber\\
\mp \frac{i}{4}\sum_{\ell=1}^{j-1} e^{\mp\alpha}a_{2j-2\ell}^{y(x)}a_{2\ell-1}^{x(y)}+e^{\pm\alpha}a_{2j-2\ell}^{y(x)}a_{2\ell+1}^{x(y)}\nonumber\\
\qquad\qquad \mp\frac{i}{4}\sum_{\ell=1}^{j-1} e^{\mp\alpha }a_{L+1-2\ell}^{y(x)}a_{L-2j+2\ell}^{x(y)}+ e^{\pm\alpha}a_{L+1-2j+2\ell}^{y(x)}a_{L+2-2\ell}^{x(y)}\, .
\end{eqnarray}
In contrast to the even case, these are not the only local conservation laws. 
Using \eref{eq:ONodd}, we can construct the following independent conserved quantities
\begin{eqnarray*}
\tilde G_{f}^{+}=i\sum_\phi f(\phi)\Bigl[\tilde b^\dag_{+}(\phi)\tilde b_-(\phi)-\tilde b^\dag_{-}(\phi)\tilde b_+(\phi)\Bigr]\\
\tilde G_{f}^{-}=\sum_\phi f(\phi)\Bigl[\tilde b^\dag_{+}(\phi)\tilde b_-(\phi)+\tilde b^\dag_{-}(\phi)\tilde b_+(\phi)\Bigr]\, ,
\end{eqnarray*}
which, in terms of the Majorana fermions, read as
\begin{eqnarray}
\fl\tilde G_{f}^{+}=2\sum_\phi\sum_{n=1}^{\frac{L+1}{2}}\sum_{\ell=1}^{\frac{L-1}{2}}\frac{f(\phi)[\sin(n\phi)e^{-\alpha}+\sin((n-1)\phi)e^{\alpha}]\sin(\ell\phi)}{(L+1)\varepsilon(\phi)}ia_{2n-1}^{x}a_{2\ell}^x\nonumber\\
+2\sum_\phi\sum_{n=1}^{\frac{L-1}{2}}\sum_{\ell=1}^{\frac{L+1}{2}}\frac{f(\phi)\sin(n\phi)[\sin(\ell\phi)e^{\alpha}+\sin((\ell-1)\phi)e^{-\alpha}]}{(L+1)\varepsilon(\phi)}ia_{2n}^{y}a_{2\ell-1}^y\label{eq:Gf+}\\
\fl\tilde G_{f}^{-}=\nn
\fl -2\sum_{n,\ell=1\atop \phi}^{\frac{L+1}{2}}\frac{f(\phi)[\sin(n\phi)e^{-\alpha}+\sin((n-1)\phi)e^{\alpha}][\sin(\ell\phi)e^{\alpha}+\sin((\ell-1)\phi)e^{-\alpha}]}{(L+1)\varepsilon^2(\phi)}ia_{2n-1}^{x}a_{2\ell-1}^y\nonumber\\
+2\sum_{n,\ell=1\atop \phi}^{\frac{L-1}{2}}\frac{f(\phi)\sin(n\phi)\sin(\ell\phi)}{L+1}ia_{2n}^{y}a_{2\ell}^x\label{eq:Gf-}
\end{eqnarray}
From these expressions we infer that a function $f(\phi)$ that makes the conservation law local (for $\alpha\neq 0$) must have the form $\varepsilon(\phi)\cos(j\phi)$, for $\tilde G_{f}^{+}$, and $\varepsilon^2(\phi)\cos(j\phi)$, for $\tilde G_{f}^{-}$. In the latter case we are using that $\varepsilon^2(\phi)$, \eref{eq:epsilon}, has only a finite number of nonzero Fourier coefficients. 

It turns out that the local conservation laws $\tilde G_{f_j}^+$ can be directly obtained from \eref{eq:tildeQodd}. Indeed they have exactly the same form as $\tilde Q^+_{f_j}+\tilde Q^-_{f_j}$ after the mapping $a_{2\ell}^x\rightarrow a_{2\ell}^y$, $a_{2\ell}^y\rightarrow -a_{2\ell}^x$. This transformation is nothing but a rotation about $z$; we indeed find
\be
\tilde G_{f_j}^+=\exp\Bigl[-i\frac{\pi}{2}\sum_{\ell=1}^{\frac{L-1}{2}}\sigma_{2\ell}^z\Bigr](\tilde Q^+_{f_j}+\tilde Q^-_{f_j})\exp\Bigl[i\frac{\pi}{2}\sum_{\ell=1}^{\frac{L-1}{2}}\sigma_{2\ell}^z\Bigr]\, .
\ee
The calculation of $\tilde G_f^-$ is more involved and tedious. Since there is no advantage in an explicit calculation, we rely on the procedure described at the beginning of \sref{s:finf} and consider directly the thermodynamic limit.

\subsection{Semi-infinite chain}

In this section we deduce the form of the symbols $\hat q_T(k)$ and $\hat q_H(k)$ associated with the local charges and defined in \eref{eq:QTH}. 
To this aim, we consider the most general term that appears in the conservation laws constructed so far, \emph{i.e.}
\be\label{eq:A}
\fl\qquad A^{\alpha\beta;j}_{s_1,s_2;s_1' s_2'}=\frac{1}{L+1}\sum_{\ell,n=1\atop\phi}\cos(j\phi)\sin((\ell-s_1)\phi)\sin((n-s_2)\phi)i a^{\alpha}_{2\ell-1+s_1'}a^\beta_{2n-1+s_2'}\, ,
\ee
where $\alpha,\beta,s_1,s_2,s_1',s_2'\in\{0,1\}$ and, for $\alpha$ and $\beta$,  $0\equiv x$ and $1\equiv y$. In the thermodynamic limit we can ignore the last term of \eref{eq:indentity} (it corresponds to the right boundary) and we have
\begin{equation*}
\fl\qquad\frac{1}{L+1}\sum_k \cos(j\phi_k)\sin(\ell\phi_k) \sin(n \phi_k)\rightarrow \frac{\delta_{\ell-n, j}+\delta_{\ell-n, -j}-\delta_{\ell+n, j}}{8}\, .
\end{equation*}
Using this relation, a generic term \eref{eq:A} of a conserved quantity has the following form
\begin{eqnarray*}
\fl\qquad A^{\alpha\beta;j}_{s_1,s_2;s_1' s_2'}\rightarrow\frac{1}{4}\sum_{n,\ell=1}^\infty\frac{\delta_{\ell-n, j-s_1+s_2}+\delta_{\ell-n, -j-s_1+s_2}-\delta_{\ell+n, j+s_1+s_2}}{4}\nonumber\\
\qquad\qquad\qquad\qquad\times (i a^{\alpha}_{2\ell-1+s_1'}a^\beta_{2n-1+s_2'}-i a^\beta_{2n-1+s_2'}a^{\alpha}_{2\ell-1+s_1'})\, .
\end{eqnarray*}
From this expression we can identify the symbols associated with this generic quadratic form of fermions, which we call $\hat a^{(s_1,s_2;j)}_T(k)$, for the Toeplitz part, and $\hat a^{(s_1,s_2;j)}_H(k)$, for the Hankel one. 
The only nonzero elements of the symbols are the following
\begin{eqnarray*}
{}[\hat a^{(s_1,s_2;j)}_T(k)]_{2s_1'+\alpha, 2s_2'+\beta}=i\frac{\cos(j k)}{2}e^{i(s_1-s_2)k}\\
{}[\hat a^{(s_1,s_2;j)}_T(k)]_{2s_2'+\beta, 2s_1'+\alpha}=[\hat a^{(s_1,s_2;j)}_T(k)]_{2s_1'+\alpha, 2s_2'+\beta}^\ast\\
{}[\hat a^{(s_1,s_2;j)}_H(k)]_{2s_1'+\alpha, 2s_2'+\beta}=i\frac{e^{ijk}}{4}e^{i(s_1+s_2-1)k}\\
{}[\hat a^{(s_1,s_2;j)}_H(k)]_{2s_2'+\beta, 2s_1'+\alpha}=-[\hat a^{(s_1,s_2;j)}_H(k)]_{2s_1'+\alpha, 2s_2'+\beta}
\end{eqnarray*}
We are now in a position to compute the symbols of the local conservation laws that are quadratic in the Majorana fermions. By inspecting the various charges, \eref{eq:Q}, \eref{eq:tildeQodd}, \eref{eq:Gf+} and \eref{eq:Gf-}, we find
\begin{eqnarray*}
\fl\qquad Q_{f_j}^{+}\sim\tilde Q_{f_j}^{+}\rightarrow 2(e^{-\alpha}A^{10;j}_{0010}+e^{\alpha}A^{10;j}_{0110})\\
\fl\qquad Q_{f_j}^{-} \sim\tilde Q_{f_j}^{-}\rightarrow-2(e^{- \alpha}A^{01;j}_{0110}+e^{\alpha}A^{01;j}_{0010})\\
\fl\qquad\tilde G^+_{f_j}\rightarrow 2e^{-\alpha}(A^{00;j}_{0001}+A^{11;j}_{0110})+2e^{\alpha}(A^{00;j}_{1001}+A^{11;j}_{0010})\\
\fl\qquad\tilde G^-_{\varepsilon f_j}\rightarrow 2e^{-2\alpha}(-A^{01;j}_{0100}+A^{10;j}_{0011})-2(A^{01;j}_{0000}+A^{01;j}_{1100})+2e^{2\alpha}(-A^{01;j}_{1000}+A^{10;j}_{0011})\nonumber\\
\qquad +2(A^{10;|j-1|}_{0011}+A^{10;j+1}_{0011})
\end{eqnarray*}
We call $\hat q^{(\pm,j)}$, $\hat {\tilde g}^{(+,j)}$, and $\hat {\tilde g}^{(-,j)}$ the symbols associates with $Q_{f_j}^{\pm}$, $-\tilde G^+_{f_j}$, and $\tilde G^-_{\varepsilon f_j}$, respectively. 
The Toeplitz part is given by
\begin{eqnarray}\label{eq:symbT}
\fl\qquad\qquad\hat q^{(+,j)}_T(k)+\hat q^{(-,j)}_T(k)=\cos(j k)\varepsilon(k)\sigma^xe^{-i\frac{k}{2}\sigma^z}\otimes\sigma^y e^{-i\theta_k\sigma^z}\nn
\fl\qquad\qquad\hat q^{(+,j)}_T(k)-\hat q^{(-,j)}_T(k)=\cos(j k)\varepsilon(k)\sigma^ye^{-i\frac{k}{2}\sigma^z}\otimes\sigma^x e^{-i\theta_k\sigma^z}\nn
\fl\qquad\qquad\hat {\tilde g}^{(+,j)}_T(k)=\cos(j k)\varepsilon(k)(\sigma^y e^{-i\frac{k}{2}\sigma^z}\otimes\sigma^z)e^{-i\theta_k\sigma^z\otimes \sigma^z}\nn
\fl\qquad\qquad\hat {\tilde g}^{(-,j)}_T(k)=
\cos(jk)\varepsilon(k) [\varepsilon(k) \1\otimes\sigma^y e^{-i\theta_k\sigma^z}]e^{-i\theta_k\sigma^z\otimes \sigma^z}\, .
\end{eqnarray}
The Hankel part (responsible for the boundary terms) reads as
\begin{eqnarray}
\fl\qquad\qquad\hat q^{(+,j)}_H(k)+\hat q^{(-,j)}_H(k)=\frac{e^{i (j-\frac{1}{2})k}}{2}\varepsilon(k)(\sigma^x\otimes\sigma^y) e^{i\theta_k\sigma^z\otimes \sigma^z}\nn
\fl\qquad\qquad\hat q^{(+,j)}_H(k)-\hat q^{(-,j)}_H(k)=\frac{e^{i (j-\frac{1}{2})k}}{2}\varepsilon(k)(\sigma^y\otimes\sigma^x) e^{i\theta_k\sigma^z\otimes \sigma^z}\nn
\fl\qquad\qquad\hat {\tilde g}^{(+,j)}_H(k)=\frac{e^{i (j-\frac{1}{2})k}}{2}\varepsilon(k)\sigma^y\otimes(\sigma^z e^{i\theta_k\sigma^z})\nn
\fl\qquad\qquad\hat {\tilde g}^{(-,j)}_H(k)=\frac{e^{i (j-\frac{1}{2})k}}{2}\varepsilon^2(k)e^{i\frac{k}{2}\sigma^z} \otimes\sigma^y
\end{eqnarray}
The dispersion relation is $\varepsilon_k=(\cos^2\frac{k}{2}+\gamma^2\sin^2\frac{k}{2})^{1/2}$ and the Bogoliubov angle $e^{i\theta_k}=(\cos \frac{k}{2}+i\gamma\sin \frac{k}{2})/\varepsilon_k$. In fact, we removed an irrelevant multiplicative constant that was reminiscent of our convention $J=2\cosh\alpha$ \eref{eq:paramXY}.  
These symbols completely characterize the local conservation laws and can also be used to reconstruct the conserved quantities in the finite chain, as explained at the beginning of \sref{s:lclXY}. 

By looking at the Toeplitz part, we recognize the conservation laws in the infinite chain that ``survive'' the chain cut (see \tref{t:charge}):
\be\label{eq:charges}
\fl\qquad  \begin{array}{rcl}
 Q^+_{f_{n}}+Q^-_{f_{n}} &\leftarrow& I_{n}^{+(e)}\\
 Q^+_{f_{n}}-Q^-_{f_{n}} &\leftarrow& J_{n}^{+(e)}\\
\tilde G^+_{f_{n}}&\leftarrow&-\cosh\alpha  (J_{n+1}^{+(o)}+J_{n}^{+(o)})+\sinh\alpha  (I_{n+1}^{-(o)}-I_{n}^{-(o)})\\
\tilde G^-_{f_{n}}&\leftarrow&\cosh\alpha(I_{n+1}^{+(o)}+I_{n}^{+(o)})-\sinh\alpha(J_{n+1}^{-(o)}-J_{n}^{-(o)})\, ,
\end{array}
\ee
where $n=0,1,\dots$\footnote{We are using slightly different notations with respect to Ref.~\cite{F:super}, where instead indices started \mbox{from $1$}.}. 
The notations for the charges in the infinite chain (r.h.s.)  are explained in \sref{s:lcl}.

\subsection{The extinct charges}\label{s:extXY}
Based on the analysis in the finite chain, one might conclude that the  local conservation laws in the periodic chain that have no analogues in the open chain are extinct:
\be\label{eq:extinction}
\begin{array}{rcl}
0&\leftarrow& I_{n}^{-(e)}\\
0&\leftarrow& J_{n}^{-(e)}\\
0&\leftarrow&\cosh\alpha  (I_{n+1}^{-(o)}+I_{n}^{-(o)})-\sinh\alpha  (J_{n+1}^{+(o)}-J_{n}^{+(o)})\\
0&\leftarrow&\cosh\alpha(J_{n+1}^{-(o)}+J_{n}^{-(o)})-\sinh\alpha(I_{n+1}^{+(o)}-I_{n}^{+(o)})\, .
\end{array}
\ee
As a matter of fact, we have not yet proven this statement, indeed we only showed that in the finite chain there are no conserved operators with such bulk part. In principle, we could be in an anomalous situation in which in the semi-infinite chain there are local conservation laws that have no analogues in finite chains (we have already seen it to happen when the number of sites is even).
We provide however some arguments, based on symmetries, in favor of \eref{eq:extinction}. 

Let us assume by contradiction that there is a local conservation law with bulk part given by a linear combination of \eref{eq:extinction}.
Since the Hamiltonian is invariant under spin flip, the charge can be always chosen to be even or odd under that transformation.  This implies that the postulated charge should be written in terms of either the first two classes of conserved quantities or the last two ones (we remind the reader that $(e)$ and $(o)$ stand for evenness and oddness under spin flip, respectively). On the other hand the charge can \emph{not} transform in itself (up to the sign) under a reflection about a bond (or about a site), otherwise we would have found it in the finite chain. This observation already rules out the existence of a local conservation law whose bulk part is a linear combination of the first two families of charges of \eref{eq:extinction}. 
The final step relies on the algebra~\cite{F:super} of the local conservation laws. Specifically, it is not possible to find a linear combination of the last two families of \eref{eq:extinction} such that the commutator with \eref{eq:charges} does not include the family $J_n^{-(e)}$. Since the commutator of two charges is a charge, this is in contradiction with our previous finding.
 
We point out that our argument relies on the finiteness of the range of the boundary term of the postulated charge. 
However, in this particular case, we do not expect that it could be possible to circumvent this contradiction relaxing locality into quasilocality. We will instead see in \sref{s:XY} that in the presence of a transverse field this possibility becomes real. 

\subsection{The XX model}\label{s:XX}
The isotropic limit $\gamma=0$, \emph{i.e.} $\alpha=0$, is somehow special.
First of all, the energy of the two families of excitations is degenerate also for chains with an even number of sites. This means that there are finite chain analogues of all the conservation laws of the semi-infinite chain. On the other hand, there is no bound state, so we lose the quasilocal conservation laws localized at the boundaries. 
There is also another peculiarity, namely the function $f$  that appears in $\tilde G^-_{f}$ is not required to be of the form $\varepsilon^2(k)\cos(j k)$, since the numerator of the first term in \eref{eq:Gf-} is factorized and one of the factors simplifies the denominator; as a consequence $\tilde G^-_{\cos(j k)}$ is local without the need of multiplying the cosine by $\varepsilon^2(k)$. In a sense, the resulting additional local conservation law (the total spin in the $z$ direction) replaces the charge that for $\alpha\neq 0$ is localized at the boundaries. We also note that adding a transverse field destroys the families of charges that are odd under a one-site shift in the bulk, \emph{i.e.} $W_j^{-_{\rm site}(o)}$ and $J_j^{+_{\rm bond}(e)}$ of \tref{t:charge}.

\subsection{Numerical analysis}
\begin{figure}
\begin{center}
\includegraphics[width=0.7\textwidth]{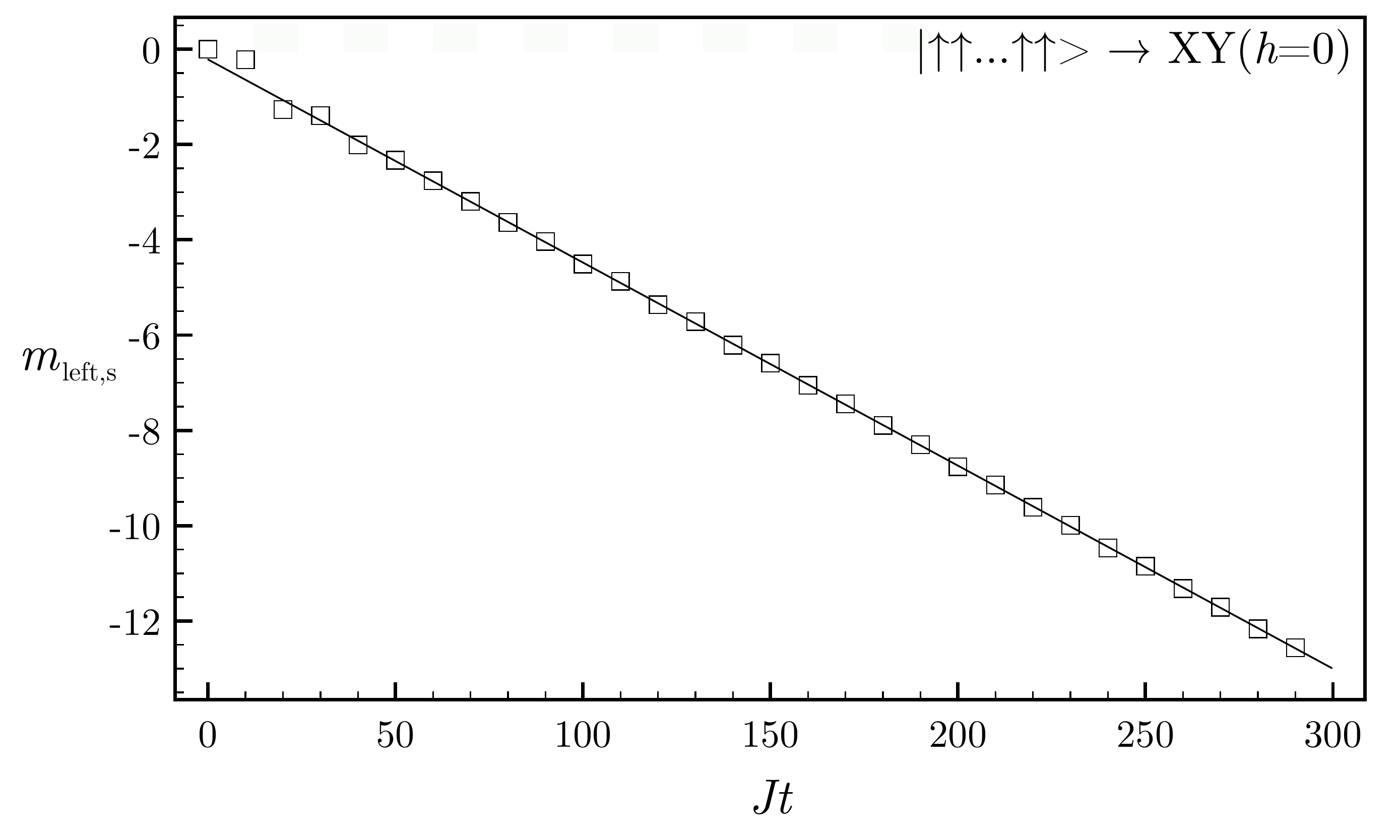}
\caption{The half-chain staggered magnetization \eref{eq:staggered} after a global quench from the state with all spins aligned along $z$ evolving under the XY Hamiltonian \eref{eq:H} with $h=0$ and $\gamma=\sqrt{2}$. The nonzero slope  is indicative of the breaking of one-site shift invariance in the bulk.}\label{f:sm}
\end{center}
\end{figure}

Remarkably, from \eref{eq:charges} and \eref{eq:extinction} it follows that the set of the local conservation laws of the semi-infinite chain does not transform well in the bulk under a shift by one site. A rather counterintuitive consequence is that one-site shift invariance can be broken evolving with the XY Hamiltonian with a boundary, without being ever restored.  

In order to check this unexpected effect, we consider the time evolution of the staggered magnetization restricted to half chain
\be\label{eq:staggered}
m_{\rm left,s}=(\lim_{L\rightarrow\infty})\frac{1}{2}\sum_{\ell=1}^{L/2}(-1)^{\ell-1}\sigma_\ell^z\, .
\ee
Since the initial state is one-site shift invariant, $m_{\rm left,s}$ is equal to zero at the initial time. 

In a chain with periodic boundary conditions shift invariance in the initial state  is sufficient for $m_{\rm left,s}$ being zero also in the limit `$\lim_{t\rightarrow\infty}\lim_{L\rightarrow\infty}$' because the local conservation laws have definite parity under a shift by one site and, in turn, the stationary state is completely characterized by the  charges that commute with the shift operator.

On the other hand, in the semi-infinite chain the charges $W$ in \tref{t:charge} do not have definite parity under a one-site shift. This implies that there are nonzero integrals of motion associated with operators that are not one-site shift invariant (in the bulk). As a consequence, despite $m_{\rm left,s}$ starting from zero, it does not approach a stationary value in the limit  `$\lim_{t\rightarrow\infty}\lim_{L\rightarrow\infty}$'; it is instead proportional to the time (see \fref{f:sm}), manifesting a light-cone propagation. 
We note that this is also sufficient to exclude the possibility that the charges $\tilde W$ in \tref{t:ext} correspond to quasilocal conservation laws in the semi-infinite chain.

This section concludes the first part of the paper, where we have computed the local conservation laws starting from the exact diagonalization of the model. In the next section the same problem will be addressed in a more abstract way and we will  consider the effects of a nonzero transverse field. 

\section{Local charges in the thermodynamic limit: a general formalism}\label{s:TH}
In \sref{s:lclTH} we introduced a correspondence between the conservation laws of a generic quadratic Hamiltonian with open boundary conditions and block-Toeplitz-plus-Hankel operators that commute with a given block-Toeplitz.
For the sake of clarity we report again the main result \eref{eq:sys}:
\be\label{eq:sys1}
\fl\qquad\qquad\left\{
\begin{array}{l}
[\hat e_{T,\kappa}(z),\hat q_{T,\kappa}(z)]=0\\
\frac{\partial^2}{\partial z^2}\Bigl[\hat e_{T,\kappa}(z)\hat q_{H,\kappa}(z)-\hat q_{H,\kappa}(z) \hat e_{T,\kappa}(1/z)\Bigr]=\Bigl[\hat q_{T,\kappa}^{+''}(0) -2\hat q_{H,\kappa}'(0)\Bigr]\hat e_{T,\kappa}^-\\
\frac{\partial^2}{\partial z^2}\Bigl[\frac{\hat q_{T,\kappa}^+(z)}{z} -\hat q_{H,\kappa}(z)\Bigr]\hat e_{T,\kappa}^-=0\\
\Bigl[\hat q_{T,\kappa}^{+''}(0)-2\hat q_{H,\kappa}'(0)\Bigr]\hat e_{T,\kappa}^- {\rm\ is\ antisymmetric}\\
\hat q_{T,\kappa}^{+'}(0)\hat e_{T,\kappa}^- +\frac{1}{2}\hat q^{''}_{H,\kappa}(0) \hat e_{T,\kappa}^++\hat q'_{H,\kappa}(0)\hat e_{T,\kappa}^0 {\rm\ is\ symmetric\, .}
\end{array}
\right.
\ee
Here $\hat e_{T,\kappa}(z)$ denotes the symbol of the block-Toeplitz operator associated with the Hamiltonian
\be
\hat e_{T,\kappa}(z)=z\hat e_{T,\kappa}^+ +\hat e_{T,\kappa}^0+\frac{1}{z}\hat e_{T,\kappa}^-
\ee
whereas $\hat q_{T,\kappa}(z)$ and $\hat q_{H,\kappa}(z)$ are the $2\kappa$-by-$2\kappa$ symbols of the Toeplitz part and of the Hankel part, respectively, of the conservation law (\emph{cf}. \eref{eq:QTH}). We are using the notation
\be
\hat q_{T,\kappa}(z)=\hat q_{T,\kappa}^+(z)+\hat q_{T,\kappa}^0+\hat q_{T,\kappa}^-(1/z)
\ee
where $\hat q_{T,\kappa}^\pm(z)$ are (matrix-)polynomials that zero at $z=0$. 
Symbols associated with quadratic forms of fermions have the properties $\hat q_{T,\kappa}(z)=-\hat q_{T,\kappa}^\ast (z)=-\hat q_{T,\kappa}^t (1/z)$ and  $\hat q_{H,\kappa}(z)=-\hat q_{H,\kappa}^\ast(z)=-q_{H,\kappa}^t(z)$.
The proof of \eref{eq:sys1} is reported in \ref{a:TH}; this section is devoted to a general analysis of the solutions of \eref{eq:sys1}. 

Our first aim is to simplify the system of equations. 
Let  us assume that, for a given $\kappa$,  there are more than $\kappa(10\kappa-1)$ solutions. It is then possible to find a linear combination of the solutions with the auxiliary properties\footnote{For this estimate we are only using that $\hat q_{T,\kappa}(z)$ and $\hat q_{H,\kappa}(z)$ are purely imaginary for $z$ real and that $\hat q_{H,\kappa}(z)$ is antisymmetric.}:
\be\label{eq:prop}
\hat q_{T,\kappa}^{+''}(0)=\hat q_{T,\kappa}^{+'}(0)=\hat q_{H,\kappa}'(0)=0\, .
\ee
Using \eref{eq:prop}, system \eref{eq:sys1} can be reduced to the following one 
\be\label{eq:redsys}
\left\{
\begin{array}{l}
[\hat e_{T,\kappa}(z),\hat q_{T,\kappa}(z)]=0\\
\Bigl(\frac{1}{z}\hat q_{T,\kappa}^+(z)-\hat q_{H,\kappa}(z)\Bigr)\hat e_{T,\kappa}^-=0\\
\hat e_{T,\kappa}(z)\hat q_{H,\kappa}(z)-\hat q_{H,\kappa}(z)\hat e_{T,\kappa}(\frac{1}{z})=0\, .
\end{array}
\right.
\ee
One can easily check that any solution of \eref{eq:redsys} is also solution of \eref{eq:sys1}. In addition, assuming \eref{eq:prop}, \eref{eq:redsys} is invariant under the transformation
\be\label{eq:transfqT}
\hat q_{T,\kappa}^+ (z)\rightarrow z^j\hat q_{T,\kappa}^+(z)\qquad \hat q_{H,\kappa} (z)\rightarrow z^j\hat q_{H,\kappa}(z)
\ee
for any integer $j\geq -1$ (but also for $j=-2$ if $\hat q^{'}_{H,\kappa}(0)=0$). The practical effect of the transformation is to  increase the range of the operator by $j \kappa$. This is therefore a \emph{systematic procedure to generate  an infinite set of local conservation laws} (which, for $j<0$, includes also one or two solutions that do not satisfy \eref{eq:prop}).

By reversing the reasoning we can state that, for $\kappa$ fixed, there can be only a finite number of linearly independent solutions of \eref{eq:sys1} which are not solutions of the reduced system~\eref{eq:redsys}.  In the following we will refer to the solutions of \eref{eq:redsys} as the ``regular solutions'' of \eref{eq:sys1} (we stress again that they do not need to satisfy \eref{eq:prop}). The remaining ones will be qualified as ``atypical''. 

\subsection{Chain inversion}\label{ss:refsym}
As pointed out in \cite{GM:open}, chain inversion plays a key role in chains with open boundary conditions. In the framework of algebraic Bethe ansatz, \cite{GM:open} showed that, in the XYZ model, all the local conservation laws which can be obtained by expanding the logarithm of the boundary transfer matrix around the shift point are invariant under chain inversion. Using a direct approach, we already showed that this rule is not free from exceptions. Specifically, if the chain is odd the family of charges $W_j^{-_{\rm site}(o)}$ turns out to be odd under that symmetry\footnote{We note that this is not in contradiction with the previous statement: we have not derived these conservation laws by taking the derivatives of the logarithm of the transfer matrix.}. 

In this section we investigate the simplifications that arise when a quadratic Hamiltonian is invariant under chain inversion, like in the quantum XY model in a transverse field. 

Chain inversion $\sigma_\ell^\alpha\rightarrow\sigma^\alpha_{L+1-\ell}$ , with $\alpha\in\{x,y,z\}$, corresponds to the following transformation on the Majorana fermions
\begin{equation}\label{eq:transfR}
a_\ell \rightarrow  (-1)^{\ell+1}i \Pi^z a_{2L+1-\ell}\, .
\end{equation}
Alternatively, for any given quadratic operator of the form \eref{eq:Qrep0} and \eref{eq:QTH}, one can apply the following transformation to the operator $\mathcal Q$:
\be\label{eq:Ref}
\mathcal Q_{2\kappa\ell+j,2\kappa\ell'+j'}\rightarrow (-1)^{j+j'}\mathcal Q_{2\kappa(L-\ell)+1-j, 2\kappa(L-\ell')+1-j'}\, .
\ee
As a result, the symbol of the Toeplitz part  transforms as follows:
\be\label{eq:cRev}
\hat q_{T,\kappa}(z)\rightarrow\Sigma_\kappa^x \otimes \sigma^y \hat q_{T,\kappa}(1/z) \Sigma_\kappa^x\otimes \sigma^y\, ,
\ee
where $\Sigma_\kappa^x$ is the $\kappa$-by-$\kappa$ exchange matrix $[\Sigma_\kappa^x]_{i j}=\delta_{i+j,\kappa+1}$.
 If a quadratic form of fermions is invariant under chain inversion, its Toeplitz symbol $\hat q_T(z)$ must be invariant under \eref{eq:cRev} for any $\kappa$.  
Being the XY Hamiltonian reflection symmetric, we then have
\be
\hat e_{T,\kappa}(1/z)=\Sigma_\kappa^x \otimes \sigma^y \hat e_{T,\kappa}(z) \Sigma_\kappa^x\otimes \sigma^y\, .
\ee
This identity allows us to recast the third equation of \eref{eq:sys1} (and \eref{eq:redsys}) in the form of a commutator:
\begin{equation*}
\hat e_{T,\kappa}(z)\hat q_{H,\kappa}(z)-\hat q_{H,\kappa}(z)\hat e_{T,\kappa}(1/z)=[\hat e_{T,\kappa}(z),\hat q_{H,\kappa}(z)\Sigma_\kappa^x\otimes \sigma^y]\Sigma_\kappa^x\otimes \sigma^y
\end{equation*}
We can therefore rewrite system \eref{eq:sys1} as follows
\be\label{eq:sys2}
\left\{
\begin{array}{l}
[\hat e_{T,\kappa}(z),\hat q_{T,\kappa}(z)]=0\\
\frac{\partial^2}{\partial z^2}\Bigl[\frac{\hat q_{T,\kappa}^+(z)}{z} -\hat q_{H,\kappa}(z)\Bigr]\hat e_{T,\kappa}^-=0\\
\frac{\partial^2}{\partial z^2}[\hat e_{T,\kappa}(z),\hat q_{H,\kappa}(z)\Sigma_\kappa^x\otimes \sigma^y]=\Bigl[\hat q_{T,\kappa}^{+''}(0) -2\hat q_{H,\kappa}'(0)\Bigr]\hat e_{T,\kappa}^-\Sigma_\kappa^x\otimes \sigma^y\\
\Bigl[\hat q_{T,\kappa}^{+''}(0)-2\hat q_{H,\kappa}'(0)\Bigr]\hat e_{T,\kappa}^- {\rm\ is\ antisymmetric}\\
\hat q_{T,\kappa}^{+'}(0)\hat e_{T,\kappa}^- +\frac{1}{2}\hat q^{''}_{H,\kappa}(0) \hat e_{T,\kappa}^++\hat q'_{H,\kappa}(0)\hat e_{T,\kappa}^0 {\rm\ is\ symmetric\, .}
\end{array}
\right.
\ee
Analogously the reduced system \eref{eq:redsys} becomes
\be\label{eq:redsys1}
\left\{
\begin{array}{l}
[\hat e_{T,\kappa}(z),\hat q_{T,\kappa}(z)]=0\\
\Bigl(\frac{1}{z}\hat q_{T,\kappa}^+(z)-\hat q_{H,\kappa}(z)\Bigr)\hat e_{T,\kappa}^-=0\\
\Bigl[\hat e_{T,\kappa}(z),\hat q_{H,\kappa}(z)\Sigma_\kappa^x\otimes \sigma^y\Bigr]= 0\, .
\end{array}
\right.
\ee

\subsection{Form of the solutions for ``generic'' noninteracting Hamiltonians}\label{s:form}
Let us consider a noninteracting Hamiltonian of a generic model that in the infinite chain has only a standard  abelian set of local conservation laws, which can be written as linear combinations of the mode occupation numbers. As shown in \cite{F:super}, this is possible only if, for generic $z$, the symbol $\hat e_{T,\kappa}(z)$ associated with the Hamiltonian is nondegenerate. 
In such generic situation the solutions of the first equation of \eref{eq:sys1}, namely
\begin{equation*}
[\hat q_{T,\kappa}(z),\hat e_{T,\kappa}(z)]=0\, ,
\end{equation*}
are simply polynomials of $\hat e_{T,\kappa}(z)$:
\be\label{eq:qTc}
\hat q_{T,\kappa}(z)=\sum_{j=0}^{2\kappa-1}\frac{(-i)^{j-1}c_j(z)+i^{j-1}c_j(1/z)}{2}[\hat e_{T,\kappa}(z)]^j\, .
\ee
Here $c_j(z)$ are polynomials of $z$ with real coefficients and we used the properties $\hat q_{T,\kappa}(z)=-\hat q_{T,\kappa}^\ast (z)=-\hat q_{T,\kappa}^t (1/z)$. 

We now focus on Hamiltonians which are invariant under chain inversion. As shown in  \sref{ss:refsym}, this implies that the Toeplitz operator $\mathcal H$ associated with the Hamiltonian is invariant under the unitary Hermitian transformation
\be
\mathcal R=\Sigma^x_{L/\kappa}\otimes \Sigma^x_{\kappa}\otimes \sigma^y
\ee
where the indices of the first operator in the Kronecker product are the same that label the blocks in \eref{eq:QTH}. 
In the most generic case, there are no degeneracies in the spectrum of $\mathcal H$ and the eigenvectors of $\mathcal H$ must be also eigenvectors of $\mathcal R$.  
As shown in \sref{ss:frep}, the excitations of the model are in a simple relation with the eigenvectors of   $\mathcal H$. We can therefore expect that under chain inversion the excitations $b^\dag(\phi)$ will transform as $b^\dag(\phi)\mapsto s(\phi) \Pi^z b^\dag(\phi)$, where $s(\phi)$ is a sign (\emph{cf.} \eref{eq:transfR}).  On the other hand, in the absence of degeneracy, any noninteracting conservation law is written in terms of the mode occupation numbers $b^\dag(\phi)b(\phi)$. Thus, \emph{it has to be even under chain inversion}. If this property is preserved in the thermodynamic limit, we can restrict even more the class of possible symbols for the Toeplitz part of a local charge:
\be\label{eq:qTc1}
\hat q_{T,\kappa}(z)=\sum_{j=0}^{\kappa-1} \frac{p_{j}(z)+p_{j}(1/z)}{2}[\hat e_{T,\kappa}(z)]^{2j+1}
\ee
where  $p_j(z)=(-1)^j c_{2j+1}(z)$ are polynomials and we can impose $p_0(0)=0$ in order to exclude the Hamiltonian. 

If we focus on the regular solutions (the solutions of \eref{eq:redsys1}) and \emph{assume invariance under chain inversion}, the absence of degeneracy in $\hat e_{T,\kappa}(z)$ allows us to fix also the form of the symbol of the Hankel part which has to satisfy the last equation of \eref{eq:redsys1}). Specifically we find
\be
\hat q_{H,\kappa}(z)=\sum_{j=0}^{\kappa-1}d_{j}(z) [\hat e_{T,\kappa}(z)]^{2j} \Sigma_\kappa^x\otimes \sigma^y
\ee
where $d_{j}(z)$ have real coefficients and are such that $d_{j}(z) [\hat e_{T,\kappa}(z)]^{2j}$ do not have negative powers of $z$.
 
The second equation of \eref{eq:redsys1} is the link between the Topelitz part and the Hankel one. Using \eref{eq:transfqT}, for regular solutions with sufficiently wide range we can assume
\begin{equation*}
\hat q_{T,\kappa}^+(z)=\frac{1}{2}\sum_{j=0}^{\kappa-1} p_{j}(z)[\hat e_{T,\kappa}(z)]^{2j+1}
\end{equation*}
and hence the second equation of \eref{eq:redsys} reads as
\be\label{eq:redsol}
\sum_{j=0}^{\kappa-1} [\hat e_{T,\kappa}(z)]^{2j}\Bigl[\frac{1}{2}\frac{p_{j}(z)}{z}\hat e_{T,\kappa}(z)-d_{j}(z) \Sigma_\kappa^x\otimes \sigma^y\Bigr]\hat e_{T,\kappa}^-=0\, .
\ee
As far as the reduced system \eref{eq:redsys1} is concerned, the problem of identifying the local conservation laws is  equivalent to finding $p_{j}(z)$ and $d_{j}(z)$ that solve \eref{eq:redsol}.

If all the reflection symmetric local conservation laws in the infinite chain have analogues in the semi-infinite chain and the boundary part of any charge can be described by a block-Hankel matrix with fixed block's size, each term of the sum in \eref{eq:redsol} must vanish separately. Indeed one can seek for solutions with $p_j(z)$ nonzero only for one given $j$ at a time. 
Consequently, the following system should have solution
\be\label{eq:simplesol}
\left\{
\begin{array}{l}
(\hat e_{T,\kappa}^++a_+\Sigma_\kappa^x\otimes \sigma^y)\hat e_{T,\kappa}^-=0\\
(\hat e_{T,\kappa}^0+a_0\Sigma_\kappa^x\otimes \sigma^y)\hat e_{T,\kappa}^-=0\\
(\hat e_{T,\kappa}^-+a_-\Sigma_\kappa^x\otimes \sigma^y)\hat e_{T,\kappa}^-=0
\end{array}
\right.
\ee
with $a_0$, $a_+$ and $a_-$ three auxiliary parameters.

\subsection{Transverse-field Ising chain}\label{s:lclTFIC}
In this section we apply the formalism developed to the transverse-field Ising chain. The reader interested in the direct calculation can find some details in \ref{a:TFIC}.

Contrary to the XY model, the TFIC does not possess non-commuting local charges and belongs to the class of models that can be described through the equations derived in \sref{s:form}.  
Let us consider the representation with $\kappa=1$. 
The symbol of the Hamiltonian is given by (\emph{cf.} \eref{eq:blockT} and \eref{eq:QTH})
\begin{equation*}
\hat e_{T,1}^+=-i\sigma^+\qquad \hat e_{T,1}^0=-h\sigma^y\, .
\end{equation*}
System \eref{eq:simplesol} reads as
\begin{equation*}
\left\{
\begin{array}{l}
(-\frac{i}{2}\sigma^++a_+ \sigma^y)\sigma^-=0\\
(-h\sigma^y+a_0\sigma^y)\sigma^-=0\\
(\frac{i}{2}\sigma^-+a_- \sigma^y)\sigma^-=0\, .
\end{array}
\right.
\end{equation*}
This can be readily solved and we find
\begin{equation*}
a_+=-1\qquad a_0=h\qquad a_-=0\, .
\end{equation*}
Coming back to \eref{eq:redsol}, this means
\begin{equation*}
d_0(z)=(z-h)\frac{{p_0(z)}}{2z}
\end{equation*}
and hence
\begin{eqnarray*}
\hat q_{T,1}(z)=\frac{p_{0}(z)+p_{0}(1/z)}{2}\hat e_{T,1}(z)\\
\hat q_{H,1}(z)=(z-h)\frac{{p_0(z)}}{2z}\sigma^y\, .
\end{eqnarray*}
By choosing $p_0(z)=z^j$ we obtain the local conservation laws reported in \tref{t:charge}. One can verify that these are solutions of system \eref{eq:sys2} also for $j=1$ (the constant term in $\hat q_H(z)$ can be dropped), showing that  any reflection symmetric charge of the TFIC in the infinite chain has an analogue in the semi-infinite chain. 

\subsection{Quantum XY model in a transverse field}\label{s:XY}
In this section we consider the quantum XY model in a transverse field with $h\neq 0$ and $\gamma\neq 1$. The local conservation laws in the infinite chain commute with one another, so we can apply again the results of \sref{s:form}.  However, we note that in the thermodynamic limit for $h<|\gamma^2-1|$ the spectrum of $\mathcal H$ becomes partially degenerate and this could invalidate our assumption that there are not charges which are odd under chain inversion. However, in this paper we ignore this complication and look only for reflection symmetric charges.

For $\kappa=1$ the symbol of the Hamiltonian is given by
\begin{equation*}
\hat e_{T,1}^+=\frac{\sigma^y-i\gamma\sigma^x}{2}\qquad \hat e_{T,1}^0=-h\sigma^y\, .
\end{equation*}
There is a very important difference with respect to the Ising case: $\hat e_{T,1}^\pm$ are invertible. This means that the second equation of \eref{eq:sys2} can be readily solved
\begin{equation*}
\hat q_{H,1}(z)=\lambda z\sigma^y -\frac{\hat q_{T,1}^+(z)}{z}+\hat q_{T,1}^{+'}(0)\, ,
\end{equation*}
where the form of the first term on the right hand side is fixed by the fact that $\hat q_{H,1}(z)$ must be antisymmetric (and purely imaginary) and hence can not be other than proportional to $\sigma^y$. 
On the other hand, in the infinite chain there is no charge besides the Hamiltonian with $\partial_z \frac{\hat q_{T,1}^+(z)}{z}\propto \sigma^y$, therefore  for $\kappa=1$ the system has no  solution. 

Let us consider $\kappa=2$. The symbol of the Hamiltonian is now given by
\begin{equation*}
\hat e_{T,2}^+=\sigma^+\otimes \frac{\sigma^y-i\gamma\sigma^x}{2}\qquad \hat e_{T,2}^0=\Bigl(\frac{\sigma^x}{2}-h \1\Bigr)\otimes\sigma^y-\gamma\frac{\sigma^y\otimes\sigma^x}{2}\, .
\end{equation*}
We note that $\hat e_T^\pm$ is not invertible anymore.
Nevertheless, system \eref{eq:redsol} has no solution. We also checked that \eref{eq:redsol} has no solution even for $\kappa=4$ and $\kappa=8$, suggesting that this model might not have regular solutions at all. 
This means that the boundary parts of the conservation laws do not have a sufficiently regular structure and we lose all the advantages of a representation in terms of Hankel operators. Nevertheless, we can still exploit the structure of the bulk part. 

Let us then seek for atypical solutions. 
We focus on a 
representation with $\kappa$ sufficiently large that  $\partial^2_z\hat q_{H,\kappa}(z)=0$. 
System \eref{eq:sys2}  becomes 
\be\label{eq:sysmax}
\fl\qquad\qquad\left\{
\begin{array}{l}
[\hat e_{T,\kappa}(z),\hat q_{T,\kappa}(z)]=0\\
\frac{\partial^3}{\partial z^3}\hat q_{T,\kappa}^+(z)\hat e_{T,\kappa}^-=0\\
\hat e_{T,\kappa}^+ \hat q_{H,\kappa}=\frac{1}{2}\hat q_{T,\kappa}^{+''}(0)\hat e_{T,\kappa}^-\\
{}[\hat e_{T,\kappa}^0,\hat q_H]=\hat q_{T,\kappa}^{+'}(0)\hat e_{T,\kappa}^--\hat e_{T,\kappa}^+\hat q_{T,\kappa}^{-'}(0)
\end{array}
\right.
\ee
Let us consider the symbol of the most general reflection symmetric conservation law. For $\kappa=1$ this has the form (we excluded the Hamiltonian)
\be\label{eq:qT}
\hat q_{T,1}(z)=\sum_{j=1}^{r} q_j \frac{z^j+z^{-j}}{2}\hat e_{T,1}(z)\, .
\ee
This expression can be used to construct the symbols for $\kappa=2^n>1$ ($n\in\mathbb N$), by means of the recursive formula~\cite{F:super}
\be\label{eq:recfor}
 \hat q_{T,\kappa}(z)=\frac{\1+\sigma^x z^{-\frac{\sigma^z}{2}}}{2}\otimes \hat q_{T,\frac{\kappa}{2}}(z^{1/2})+\frac{\1-\sigma^x z^{-\frac{\sigma^z}{2}}}{2}\otimes \hat q_{T,\frac{\kappa}{2}}(-z^{\frac{1}{2}})\, .
\ee
Let us focus on $\kappa=2$. The symbol reads as
\begin{equation*}
\fl \hat q_{T,2}(z)=\sum_{j=1}^{r}q_j \frac{z^{\frac{j}{2}}+z^{-\frac{j}{2}}}{2} \Bigl[\frac{\1+\sigma^x z^{-\frac{\sigma^z}{2}}}{2}\otimes \hat e_{T,1}(z^{1/2})+(-1)^j\frac{\1-\sigma^x z^{-\frac{\sigma^z}{2}}}{2}\otimes \hat e_{T,1}(-z^{\frac{1}{2}})\Bigr]
\end{equation*}
We find that the second equation of \eref{eq:sysmax} implies
\begin{equation*}
q_j=0\qquad \forall j\geq 5\, .
\end{equation*}
That is to say, the range of the generic charge is in fact bounded: 
\begin{equation*}
\fl \hat q_{T,2}(z)=\sum_{j=1}^{4}q_j \frac{z^{\frac{j}{2}}+z^{-\frac{j}{2}}}{2} \Bigl[\frac{\1+\sigma^x z^{-\frac{\sigma^z}{2}}}{2}\otimes \hat e_{T,1}(z^{1/2})+(-1)^j\frac{\1-\sigma^x z^{-\frac{\sigma^z}{2}}}{2}\otimes \hat e_{T,1}(-z^{\frac{1}{2}})\Bigr]
\end{equation*}
Being skew-symmetric and purely imaginary, $\hat q_H $ has the following form:
\begin{equation*}
\hat q_{H,2}=(\lambda_0 \1+\lambda_x \sigma^x+\lambda_z \sigma^z)\otimes\sigma^y+\sigma^y\otimes (\mu_0 \1+\mu_x \sigma^x+\mu_z \sigma^z)
\end{equation*}
where all the parameters are real. In conclusion, if the system of equations \eref{eq:sysmax} has a solution, it should be possible to find ten real parameters $\{q_1,q_2,q_3,q_4,\lambda_0,\lambda_x,\lambda_z,\mu_0,\mu_x,\mu_z\}$ satisfying the last two equations of \eref{eq:sysmax}. 
Indeed, we find a solution (up to a multiplicative factor):
\begin{eqnarray*}
\fl\qquad\qquad\{q_1,q_2,q_3,q_4,\lambda_0,\lambda_x,\lambda_z,\mu_0,\mu_x,\mu_z\}\nn
=\Bigl\{4 h,\gamma^2-1,0,0,h\frac{3+\gamma^2}{4},\frac{\gamma^2-1}{4},h\frac{3+\gamma^2}{4},0,\frac{\gamma(\gamma^2-1)}{4},0\Bigr\}\, .
\end{eqnarray*}
We checked that this is the conservation law reported in Ref.~\cite{GM:open}. Up to a a term proportional to the Hamiltonian, this corresponds to multiplying $\hat e_{T,1}(z)$ in \eref{eq:qT} by
\begin{equation*}
\Bigl(\frac{z+z^{-1}}{2}+\frac{h}{\gamma^2-1}\Bigr)^2\, .
\end{equation*}

For larger $\kappa$, we expect that the second equation of \eref{eq:sysmax} gives the constraint $r=2\kappa$ (although \emph{a posteriori} we find that one could also impose $r=\kappa$). As long as $\kappa$ is a power of $2$, a rather crude procedure to solve the system of equations could be to express $\hat q_H$ as the most general purely imaginary skew-symmetric $(2\kappa)$-by-$(2\kappa)$ matrix and to manage the Toeplitz part by means of \eref{eq:recfor}.
This gives a total of $\kappa(2\kappa+1)$ parameters that must be fixed imposing the last two equations of \eref{eq:sysmax}.
For $\kappa=4$ we find one additional independent solution, whose bulk part has the symbol \eref{eq:qT} with
\be
q_1=1-\frac{8h^2}{(\gamma^2-1)^2}\qquad q_3=1\qquad q_4=\frac{\gamma^2-1}{8 h}\, .
\ee
Interestingly, up to terms  proportional to the previous charge and to the Hamiltonian, this corresponds to multiplying $\hat e_{T,1}(z)$ in \eref{eq:qT} by  
\be
\Bigl(\frac{z+z^{-1}}{2}+\frac{h}{\gamma^2-1}\Bigr)^4\, .
\ee
We solved the system of equations also for $\kappa=8$ and found two additional independent solutions of the same form but with exponent $6$ and $8$. This suggests that, for generic $\kappa$, the bulk part of the conservation law is simply given by
\be\label{eq:conjecture}
\hat q_{2j;T,1}(z)=\Bigl(\frac{z+z^{-1}}{2}+\frac{h}{\gamma^2-1}\Bigr)^{2j}\hat e_{T,1}(z)\, .
\ee
We note that the presence of only even powers of $c_{q}(z)=\frac{z+z^{-1}}{2}+q$, with $q=\frac{h}{\gamma^2-1}$, is not sufficient to conclude that the charges with PBC associated with odd powers of $c_{q}(z)$ are destroyed by the boundary. 
For example, for $|q|>1$ we can easily show that the operators corresponding to odd powers are linear combinations of the charges associated with even powers. This can be seen as follows.
For $z$ in the unit circle, being $\sqrt{(1+|q|)^2-x}$ an analytic function of $x$ in $x<(1+|q|)^2$, we can series expand $c_q(z)$ in powers of $c_q^2(z)-(1+|q|)^2$
\begin{equation*}
c_q(z)=\sum_j a_j \Bigl[\Bigl(\frac{z+z^{-1}}{2}+q\Bigr)^2-(1+|q|)^2\Bigr]^j\, ,
\end{equation*}
where the coefficients $a_j$ decay as $\sim \frac{(1+|q|)^{-2j}}{j^{3/2}}$. 
The locality properties of the charge close to the boundary can be inferred without knowing the explicit expression of $q_{H,\kappa}(z)$. First, we observe that the range of the edge part corresponding to $c_q^{2j}(z)$ is bounded from above by $2j+2$ (this is because we chose $k$ so large that the block-Hankel operator has just one block-element different from zero). Second, we normalize the argument of the expansion above in such a way that, in the unit  circle ($|z|=1$), the argument is always smaller than or equal to $1$. This implies that the corresponding charge $Q_j$ satisfies $\parallel Q_j\parallel\leq \parallel H\parallel$, where $\parallel\cdot\parallel$ is the maximal eigenvalue in absolute value (this is because the energy of any excitation of the charge is smaller than the energy of the corresponding excitation of the Hamiltonian).
Consequently, the coefficients of the expansion decay exponentially as $[1-(\frac{|q|-1}{|q|+1})^2]^j$ and the edge part of the conservation law represented by $c_q(z)$ turns out to be quasilocal. The same argument applies also to $c^{2j-1}_q(z)$ for any $j>0$. 
In conclusion, for $|q|>1$ the set \eref{eq:conjecture} is complete and all the local conservation laws that seem to be missed, not being written as a finite linear combination of \eref{eq:conjecture}, have in fact quasilocalized contributions at the boundaries. 

For $|q|<1$ the situation is different, indeed the functions $c_q^{2j}(e^{i k})$
are not complete in $L^2[0,\pi]$\footnote{We consider $[0,\pi]$ because reflection symmetry implies that the functions must be even in $[-\pi,\pi]$}. For example $f_m(k)=\sin k\ c_q^{2m+1}(e^{i k})\theta(1-2q-\cos k)$ are orthogonal to $c_q^{2j}(e^{i k})$, for any $j,m\geq 0$ (we are assuming $q>0$). On the other hand, $c_q(e^{i k})$ is not orthogonal to $f_m(k)$:
\begin{equation*}
\int_{\arccos(1-2q)}^\pi\frac{\mathrm d k}{\pi}(\cos k+q) [\sin k(\cos k+q)^{2m+1}]=\frac{2}{\pi}\frac{(1-q)^{2m+3}}{2m+3}
\end{equation*}
and hence the associated operator can not be written as a linear combination of the operators with symbols \eref{eq:conjecture}.

This is not yet sufficient to exclude the existence of quasilocal conservation laws whose Toeplitz symbols are linearly independent of \eref{eq:conjecture}. 
In the next section we provide some numerical evidence that for $h<|\gamma^2-1|$ the missing reflection symmetric charges are destroyed by the boundary, at least separately. However, we note that there are indirect numerical evidences of additional quasilocal conservation laws which are odd under chain inversion. 

\subsection{Numerical analysis}\label{s:XYhanalysis}

\begin{figure}
\begin{center}
\includegraphics[width=0.7\textwidth]{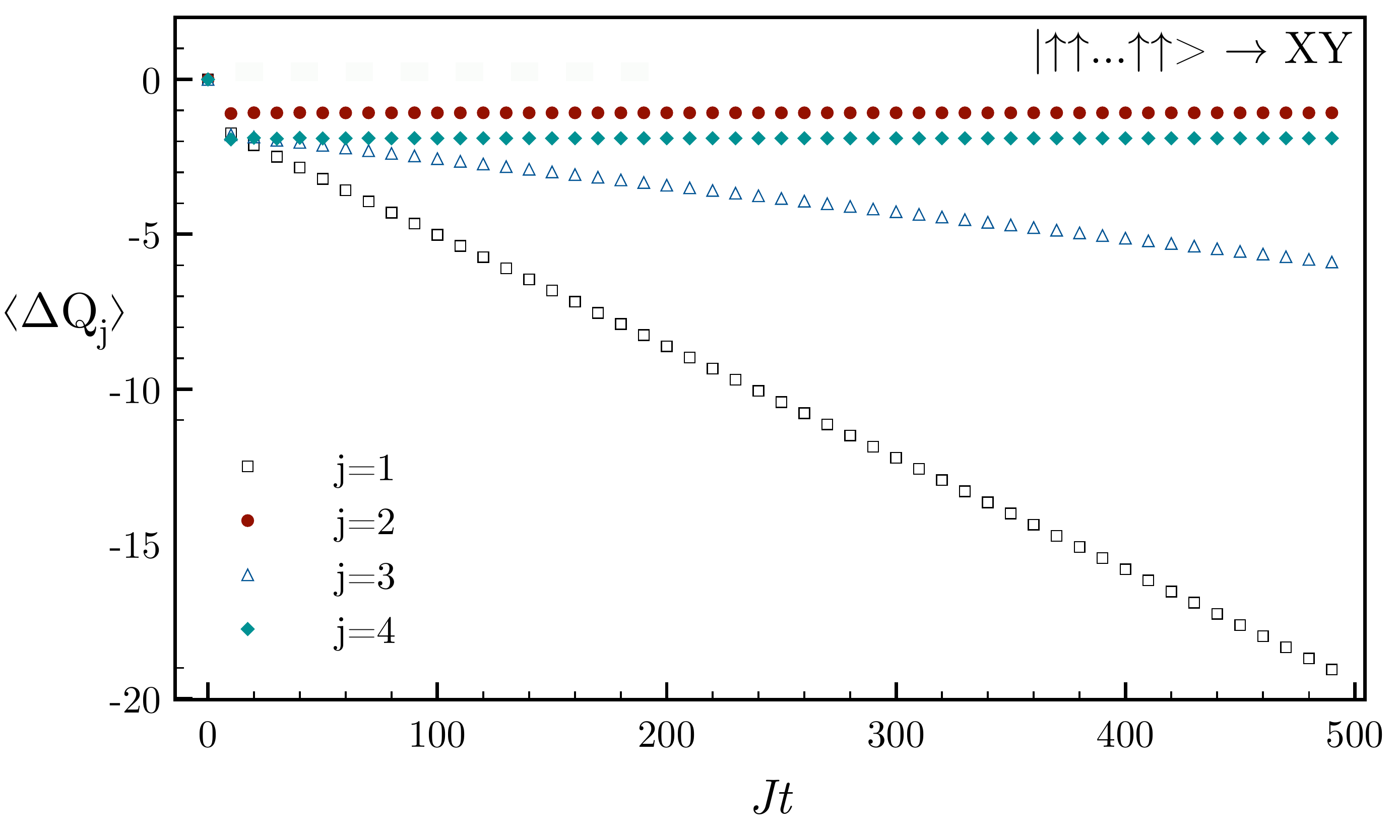}
\caption{The time evolution of $\braket{\Delta Q_j}$ \eref{eq:DeltaQj} after a global quench from the state with all spins aligned along $z$ evolving under the XY Hamiltonian \eref{eq:H} with $h=0.75$ and $\gamma=\sqrt{3}$ (thus $\frac{h}{\gamma^2-1}<1$) for various values of $j$ (we omitted $j=0$, being the energy difference and hence identically zero at any time). For $j$ even, $\braket{\Delta Q_j}$ approaches a stationary value; for $j$ odd, there is a contribution proportional to the time.}\label{f:Qn}
\end{center}
\end{figure}
\begin{figure}
\begin{center}
\includegraphics[width=0.7\textwidth]{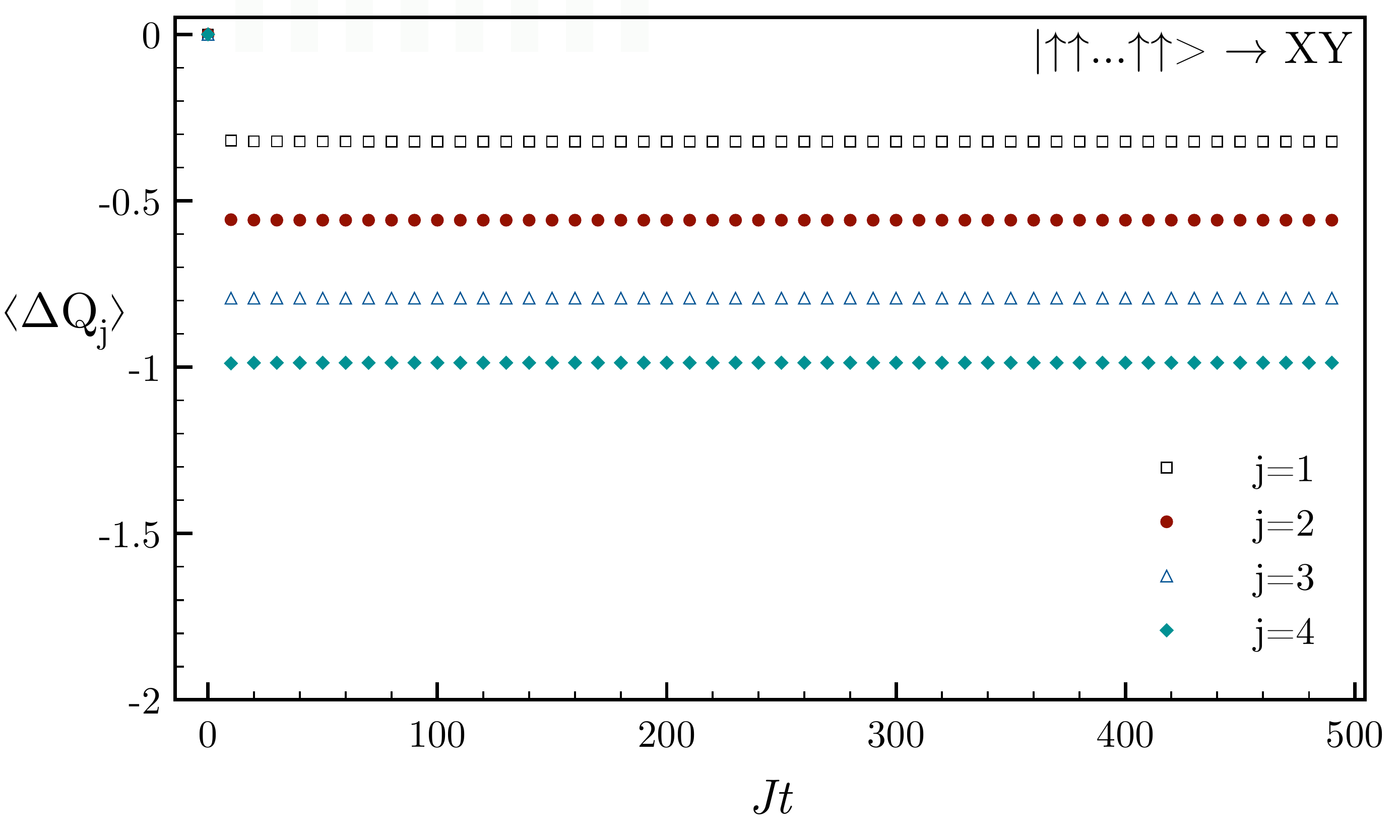}
\caption{The same as in \fref{f:Qn} for $h=2.5$, \emph{i.e.} $\frac{h}{\gamma^2-1}>1$. Independently of $j$, $\braket{\Delta Q_j}$ approaches a stationary value. }\label{f:Qnstandard}
\end{center}
\end{figure}

\begin{figure}
\begin{center}
\includegraphics[width=0.7\textwidth]{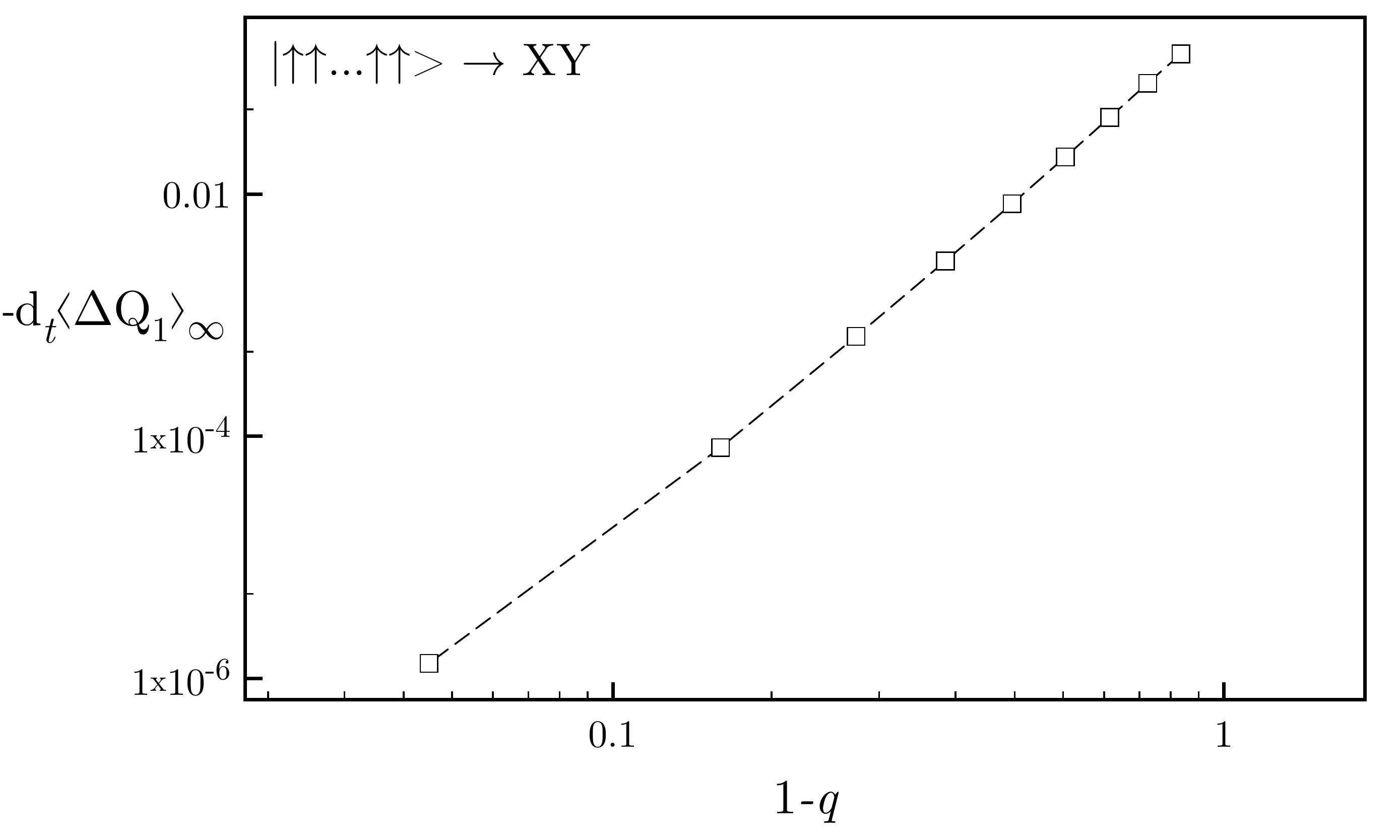}
\caption{The coefficient of the linear term in $t$ in $\braket{\Delta Q_1}$ \eref{eq:DeltaQj}  for the quench from the state with all spins aligned along $z$ evolving under the XY Hamiltonian \eref{eq:H} with $\gamma=\sqrt{3}$ and various values of  $q=\frac{h}{\gamma^2-1}$. The data are consistent with the coefficient approaching zero as a power law of $1-q$ as $q\rightarrow 1^-$. }\label{f:Q1coeff}
\end{center}
\end{figure}

In this section we consider the time evolution of the expectation values of the local quadratic forms of fermions $Q_{j}$ associated with Toeplitz operators with symbols of the form
\be\label{eq:symbtest}
\hat q_{j;T,1}(z)=\Bigl(\frac{z+z^{-1}}{2}+\frac{h}{\gamma^2-1}\Bigr)^{j}\hat e_{T,1}(z)\, .
\ee
These operators commute with the Hamiltonian up to terms localized at the boundaries. 
In \sref{s:XY} we conjectured that $Q_{2j}$ correspond to local conservation laws  in the semi-infinite chain. On the other hand, we showed that $Q_{2j-1}$ are associated with quasilocal conservation laws with OBC only for $h>|\gamma^2-1|$. In order to see what happens when $h<|\gamma^2-1|$, we consider the following quantity 
\be\label{eq:DeltaQj}
\braket{\Delta  Q_j}=(\lim_{L\rightarrow\infty}) \braket{\uparrow\cdots\uparrow|(e^{i H t}Q_je^{-i H t}-Q_j)|\uparrow\cdots\uparrow}\, .
\ee
If there is a conserved operator $\tilde Q_j$ with the same bulk part of $Q_j$ and quasilocalized contributions at the boundaries, this expectation value remains finite at any time, indeed its absolute value can be bounded from above as follows
\bea
\fl
|\braket{\uparrow\cdots\uparrow|(e^{i H t}Q_je^{-i H t}-Q_j)|\uparrow\cdots\uparrow}|\nn
\fl\qquad \leq  |\braket{\uparrow\cdots\uparrow|e^{i H t}(Q_j-\tilde Q_j)e^{-i H t}|\uparrow\cdots\uparrow}|+|\braket{\uparrow\cdots\uparrow|(\tilde Q_j-Q_j)|\uparrow\cdots\uparrow}|\nn
\leq 2\parallel \tilde Q_j-Q_j\parallel<\infty\, ,
\eea
where $\parallel\cdot\parallel$ is the operator norm. 
The last term is finite  because $\tilde Q_j-Q_j$ is assumed to be quasilocalized at the boundaries. On the other hand, if there is no boundary term that fixes the commutator with the Hamiltonian, the expectation value can diverge, but no faster then linearly in time:
\bea
\fl |\frac{\partial}{\partial t}\braket{\uparrow\cdots\uparrow|(e^{i H t}Q_je^{-i H t}-Q_j)|\uparrow\cdots\uparrow}|=|\braket{\uparrow\cdots\uparrow|[H,e^{i H t}Q_je^{-i H t}]|\uparrow\cdots\uparrow}|\nn
\leq \parallel [H,Q]\parallel<\infty\, .
\eea
The last inequality is a consequence of $Q$ being conserved in the bulk and hence the commutator with the Hamiltonian being localized at the boundaries. 
The data shown in \Fref{f:Qn} and  \Fref{f:Qnstandard}  are consistent with the following conjecture:
\begin{itemize}
\item[-] For $|h|>|\gamma^2-1|$ all the reflection symmetric conservation laws of the chain with PBC have analogues in the chain with OBC; 
\item[-] For $|h|<|\gamma^2-1|$ there is no conserved operator with bulk part described by a symbols of the form \eref{eq:symbtest} with an odd exponent. 
Nevertheless, there are still quasilocal conserved quantities written also in terms of the (individually extinct) charges corresponding to odd exponents. However, they are not enough to span the entire space of the reflection symmetric conservation laws with PBC.
\end{itemize}
\Fref{f:Q1coeff} shows that the coefficient of the linear term in $t$ in the expectation value of $Q_1$ is different from zero as long as $q=\frac{h}{\gamma^2-1}<1$ and apparently approaches zero as a power law when $q\rightarrow 1^-$.  

\section{Conclusions}\label{s:conc}
In this work we have constructed the local conservation laws of the quantum XY model with open boundary conditions. 
We used two methods: a direct approach that relies on exact diagonalization and a novel method based on the algebra of Toeplits+Hankel operators. The former is applicable to models where the excitations are sufficiently simple that one can easily guess the form of the local conservation laws in terms of the excitations. We successfully applied this method both to the quantum XY model in zero field and to the transverse-field Ising chain (\emph{cf.} \ref{a:TFIC}). The more abstract procedure based on the algebra of Toeplits+Hankel operators is generally more effective than the direct approach and
can  also be applied when the diagonalization is more complicated, as in the quantum XY model in a transverse field (\emph{cf.} \ref{a:XYh}).  

We classified the local and quasilocal conservation laws of the quantum XY model without a field.
In the thermodynamic limit we found a non-abelian set of charges. We identified conservation laws that break chain inversion and also families of local charges that, in the bulk,  do not transform well under a shift by one site. 

In the transverse-field Ising chain we obtained the expected result: all the conservation laws of the periodic chain that are odd under chain inversion disappear in the chain with open boundary conditions and any reflection symmetric charge of the infinite chain survives the boundary.

We also carried out a preliminary analysis of the quantum XY model in a transverse field.
Our main result is a conjecture about the bulk part of the (quasi)local charges. 
Remarkably, we found that the set of (quasi)local conservation laws changes crossing the curve $h=|\gamma^2-1|$, which was previously associated with a far-from-equilibrium phase transition in an open XY chain~\cite{ProsenXY}. It would be interesting to investigate whether the two phenomena are connected.
We point out that, for $|h|<|\gamma^2-1|$, a part of the energy spectrum becomes degenerate in the thermodynamic limit and the argument that we gave in \sref{s:form} against the presence of charges that are odd under chain inversion can be in fact circumvented. This leaves open the question of whether there are further, possibly odd, quasilocal charges in this region of the parameter space. 

As a by product of our investigations, we derived the conditions that the symbol of a block-Toeplitz-plus-Hankel operator must satisfy in order to commute with a block-Toeplitz. To the best of our knowledge, these conditions were never been pointed out before.

\section*{Acknowledgments}
I thank Alexander Its and Estelle Basor for correspondence. I also thank Toma\v{z} Prozen for pointing out some analogies to the results of Ref.~\cite{ProsenXY}.
This work was supported by the LabEx ENS-ICFP: ANR-10-LABX-0010/ANR- 10-IDEX-0001- 02 PSL*.

\appendix

\section[\hspace{2cm}: Toeplitz+Hankel commuting with Toeplitz]{Toeplitz+Hankel commuting with Toeplitz}\label{a:TH}
In this appendix we derive the conditions that the symbols of a block-Toeplitz+Hankel operator \eref{eq:QTH}  satisfy in order to commute with a given block-Toeplitz.
We will focus on the class of operators that are relevant to the description of noninteracting models with local interactions. 
These operators appear when Hermitian noninteracting operators are written in the following form:
\be
\hat O=\frac{1}{4}\sum_{\ell,n}a_\ell \mathcal O_{\ell n} a_n\, ,
\ee
where $a_j$ are Majorana fermions ($\{a_\ell,a_n\}=2\delta_{\ell n}$) and $\mathcal O_{\ell \ell}=0$ (the diagonal elements would give contributions proportional to the identity). The hermiticity of $\hat O$ implies the hermiticity of $\mathcal O$ and the anticommutation relations of the fermions allow us to ask for $\mathcal O$ being skewsymmetric. In conclusion, $\mathcal O$ is a purely imaginary antisymmetric operator. 
If $\hat O$ is shift invariant in the bulk, we can always isolate a block-Toeplitz part $T[\hat \phi_{\kappa}(z)]$ from $\mathcal O$, where $\hat \phi_{\kappa}(z)$ is the so-called symbol
\be
\{T[\hat \phi_\kappa(z)]\}_{2\kappa \ell+j, 2\kappa \ell' +j'}=\int_{-\pi}^\pi\frac{\mathrm d k}{2\pi}e^{-i(\ell-\ell')k}[\hat \phi_\kappa(e^{i k})]_{j j'}\, ,
\ee 
with $1\leq j,j'\leq 2\kappa$ and $\ell,\ell' \geq 0$.
If the homogeneity of $\hat O$ is broken only in a region localized at the boundary,  the rest of the operator $T[\hat \phi_\kappa(z)]-\mathcal O$ can be assumed to have block-Hankel form $H[\hat\psi_\kappa(z)]$ 
\be
\{H[\hat \psi_\kappa(z)]\}_{2\kappa \ell+j, 2\kappa \ell' +j'}=\int_{-\pi}^\pi\frac{\mathrm d k}{2\pi}e^{-i(\ell'+\ell+1)k}[\hat \psi_\kappa(e^{ik})]_{j j'}\, ,
\ee
where $\hat\psi_\kappa(z)$ is the symbol of the Hankel part. 
By writing $\mathcal O=T[\hat \phi_{\kappa}(z)]-H[\hat \psi_{\kappa}(z)]$ and imposing that $\mathcal O$ is a purely imaginary antisymmetric operator, we obtain
the conditions $\hat \phi_{\kappa}(z)=-\hat \phi_{\kappa}^\ast (z)=-\hat \phi_{\kappa}^t (1/z)$ and  $\hat \psi_{\kappa}(z)=-\hat \psi_{\kappa}^\ast(z)=-\psi_{\kappa}^t(z)$.
In addition, if $O$ is a local operator, $\mathcal O$ can have only a finite number of diagonals different from zero (around the main one). As a result, \emph{locality} implies that the symbols  must have a \emph{finite number of nonzero Fourier coefficients}. Analogously, quasilocality means that the Fourier coefficients decay exponentially.

We remind the reader of the following well-known identities that relate block-Toeplitz and block-Hankel operators~\cite{THcomm}: 
\begin{eqnarray}\label{eq:identity}
T[\hat \phi(z)\hat \psi(z)]=T[\hat \phi(z)]T[\hat \psi(z)]+H[\hat\phi(z)]H[\hat\psi(1/z)]\nonumber\\
H[\hat\phi(z)\hat\psi(z)]=T[\hat \phi(z)] H[\hat\psi(z)]+H[\hat\phi(z)]T[\hat\psi(1/z)]\, .
\end{eqnarray}
From the first equation of \eref{eq:identity} we find
\begin{equation*}
\fl \qquad [T[\hat \phi(z)],T[\hat \psi(z)]]=T[[\hat \phi(z),\hat \psi(z)]]-H[\hat\phi(z)]H[\hat\psi(1/z)]+H[\hat\psi(z)]H[\hat\phi(1/z)]
\end{equation*}
Let us apply this identity to our problem. Imposing zero commutator between the block-Toeplitz operator of the Hamiltonian $\mathcal H=T[\hat e_{T}(z)]$ and the block-Toeplitz-plus-Hankel operator associated with the local charge $\mathcal Q=T[\hat q_{T}(z)]-H[\hat q_{H}(z)]$ we obtain
\begin{eqnarray}\label{eq:commutator}
\fl\qquad 0=\Bigl[T[\hat e_T(z)],T[\hat q_T(z)]-H[\hat q_H(z)]\Bigr]=T[[\hat e_T(z),\hat q_T(z)]]\nn
\fl\qquad\qquad -H[\hat e_T(z)]H[\hat q_T(1/z)]+H[\hat q_T(z)]H[\hat e_T(1/z)]-\Bigl[T[\hat e_T(z)],H[\hat q_H(z)]\Bigr]\, .
\end{eqnarray}
Because of locality, the Toeplitz operator on the right hand side of the first line of \eref{eq:commutator} must be zero, being the only contribution in the bulk. This implies that the symbols of the Toeplitz part of the operators must commute $[\hat e_T(z),\hat q_T(z)]$.

Our goal is to express also the vanishing of the second line as a condition on the symbols. 
Let us rewrite the equation explicitly:
\begin{eqnarray}\label{eq:intTH}
\fl 0=\sum_{j=0}^\infty \int_{-\pi}^\pi\frac{\mathrm d k}{2\pi}\int_{-\pi}^\pi\frac{\mathrm d p}{2\pi}e^{-i\ell k-i n p}\Bigl[e^{i j k-i(j+1)p}\hat e_T(e^{i k})\hat q_H(e^{i p})-e^{-i (j+1) k+i j p}\hat q_H(e^{i k})\hat e_T(e^{-i p})\nn
-e^{-i (j+1)(k+p)}\Bigl(\hat q_T(e^{i k}) \hat e_T(e^{-i p})-\hat e_T(e^{i k})\hat q_T(e^{-i p})\Bigr)\Bigr]\, .
\end{eqnarray}
In the integral over $p$ of the second term we have reversed the integration variable  in order to factorize the term $e^{-i\ell k-i n p}$. This equation has the following meaning: for any $j\geq 0$, the term in square brackets has only negative Fourier coefficients; that is to say, the matrix
\be\label{eq:pol}
\fl \quad z^j\bar w^{j+1}\hat e_T(z)\hat q_H(w)-\bar z^{j+1} w^j \hat q_H(z)\hat e_T(\bar w)-\bar z^{j+1}\bar w^{j+1}\Bigl(\hat q_T^+(z) \hat e_T^-(\bar w)-\hat e_T^+(z)\hat q_T^-(\bar w)\Bigr)
\ee
can not have terms with nonnegative powers of  $z$ and $w$ simultaneously, where we used the notation $\bar z=1/z$. 
In order to write this condition in a more compact form, we consider a block dimension $\kappa$ such that 
\be\label{eq:energys}
\hat e_T(z)=z\hat e_T^++\hat e_T^0+\frac{1}{z}\hat e_T^-\, .
\ee
We then sum \eref{eq:pol} over $j$ taking only the terms with nonnegative powers of $z$ and $w$. For example, let us consider the first term of \eref{eq:pol} and \eref{eq:energys}:
\begin{eqnarray*}
\fl \sum_{j=0}^\infty z^{j+1}\bar w^{j+1}\hat e_T^+\hat q_H(w)=\sum_{j=0}^\infty \sum_{i=1} \hat e_T^+ \frac{q^{(i)}_H(0)}{i!} z^{j+1}\bar w^{j+1-i}\rightarrow   \sum_{i=1} \sum_{j=0}^{i-1}\hat e_T^+ \frac{q^{(i)}_H(0)}{i!} \Bigl(\frac{z}{w}\Bigr)^{j+1}w^{i}\nn
=\sum_{i=1}z \hat e_T^+ \frac{q^{(i)}_H(0)}{i!} \frac{w^i-z^i}{w-z}=\frac{z \hat e_T^+[\hat q_H(w)-\hat q_H(z)]}{w-z}\, .
\end{eqnarray*}
Making analogous manipulations we can recast \eref{eq:intTH}  in the following form
\begin{eqnarray*}
\fl \quad z \hat e_T^+\hat q(w)+\hat e_T^0\hat q(w)+\hat e_T^-\frac{\hat q(w)}{w} -w \hat q(w) \hat e_T^--\hat q(w)\hat e_T^0-\frac{\hat q(w)}{w}\hat e_T^+-\Bigl(\frac{z}{w}-1\Bigr)\hat e_T^+\hat q_T^-(w)\nn
\fl =z \hat e_T^+\hat q(z)+\hat e_T^0\hat q(z)+\hat e_T^-\frac{\hat q(z)}{z}-\hat q(z)w \hat e_T^--\hat q(z)\hat e_T^0-\frac{\hat q(z)}{z}\hat e_T^+-\Bigl(1-\frac{w}{z}\Bigr) \hat q_T^+(z) \hat e_T^-\, ,
\end{eqnarray*}
where, for the sake of compactness, we used the notation $\hat q(z)$ to indicate the symbol $\hat q_H(z)$ of the Hankel part. This equation should be fulfilled for any $z$ and $w$. In fact, the first (second) line is linear in $z$ ($w$). Thus, it can be simplified further. 
In particular it can be recast in a system of two equations, one for $w=0$ and one for the derivative with respect to $w$:
\begin{eqnarray*}
\fl \hat e_T^-\hat q'_0 -\hat q'_0\hat e_T^+-z\hat e_T^+\hat q_T^{-'}(0)=z \hat e_T^+\hat q(z)+\hat e_T^0\hat q(z)+\hat e_T^-\frac{\hat q(z)}{z}-\hat q(z)\hat e_T^0-\frac{\hat q(z)}{z}\hat e_T^+-\hat q_T^+(z) \hat e_T^-\nn
\fl \frac{\partial}{\partial w}\Bigl[z \hat e_T^+\hat q(w)+\hat e_T^0\hat q(w)+\hat e_T^-\frac{\hat q(w)}{w} -w \hat q(w) \hat e_T^--\hat q(w)\hat e_T^0-\frac{\hat q(w)}{w}\hat e_T^+-\Bigl(\frac{z}{w}-1\Bigr)\hat e_T^+\hat q_T^-(w)\Bigr]\nn
=-\hat q(z)\hat e_T^-+\frac{\hat q_T^+(z)}{z} \hat e_T^-
\end{eqnarray*}
The left hand side of the second equation is linear in $z$, thus, as before, can be rewritten as a system of two equations, one for $z=0$ and one for the derivative with respect to~$z$:
\begin{eqnarray*}
\fl \hat e_T(z)\hat q(z)-\hat q(z)\hat e_T(1/z)=\hat e_T^-\hat q'_0 -\hat q'_0\hat e_T^+-z\hat e_T^+\hat q_T^{-'}(0)+z\Bigl[\frac{\hat q_T^+(z)}{z}-\hat q(z)\Bigr] \hat e_T^-\nn
\fl \frac{\partial}{\partial w}\Bigl[\hat e_T^0\hat q(w)+\hat e_T^-\frac{\hat q(w)}{w} -w \hat q(w) \hat e_T^--\hat q(w)\hat e_T^0-\frac{\hat q(w)}{w}\hat e_T^++\hat e_T^+\hat q_T^-(w)\Bigr]=\hat q_T^{+'}(0)\hat e_T^-\nn
\fl \hat e_T^+ \frac{\partial}{\partial w}\Bigl[\hat q(w)-\frac{\hat q_T^-(w)}{w}\Bigr]= \frac{\partial}{\partial z}\Bigl[\frac{\hat q_T^+(z)}{z} -\hat q(z)\Bigr]\hat e_T^-=\Bigl[\frac{\hat q_T^{+''}(0)}{2} -\hat q'(0)\Bigr]\hat e_T^- \, .
\end{eqnarray*}
In the last equation we used that the two members depend on different variables, \emph{i.e.} they must be constant.
We use the third equation to remove $\hat q_T^-(w)$ from the second equation. The second equation can then be rewritten as follows:
\begin{equation*}
\fl \frac{\partial}{\partial w}\Bigl[\hat e_T(w)\hat q(w)-\hat q(w) \hat e_T(1/w)\Bigr]=-\hat e_T^+\hat q_T^{-'}(0)+\hat q_T^{+'}(0)\hat e_T^-+2w\Bigl[\frac{\hat q_T^{+''}(0)}{2} -\hat q'(0)\Bigr]\hat e_T^-\, .
\end{equation*}
Including also the equation for the Topelitz part, we finally obtain the system of equations \eref{eq:sys}
\be
\fl\qquad\qquad\left\{
\begin{array}{l}
[\hat e_T(z),\hat q_T(z)]=0\\
\frac{\partial^2}{\partial z^2}\Bigl[\hat e_T(z)\hat q_H(z)-\hat q_H(z) \hat e_T(1/z)\Bigr]=\Bigl[\hat q_T^{+''}(0) -2\hat q_H'(0)\Bigr]\hat e_T^-\\
\frac{\partial^2}{\partial z^2}\Bigl[\frac{\hat q_T^+(z)}{z} -\hat q_H(z)\Bigr]\hat e_T^-=0\\
\Bigl[\hat q_T^{+''}(0)-2\hat q_H'(0)\Bigr]\hat e_T^- {\rm\ is\ antisymmetric}\\
\hat q_T^{+'}(0)\hat e_T^- +\frac{1}{2}\hat q^{''}_H(0) \hat e_T^++\hat q'_H(0)\hat e_T^0 {\rm\ is\ symmetric\, .}
\end{array}
\right.
\ee
We note that the last two conditions rely on the transformation rules of the symbols under transposition. 

\section[\hspace{2cm}: Direct method: exact diagonalization]{Exact diagonalization}\label{A:direct}
 \subsection[\hspace{1.5cm}: XY model in zero field]{Quantum XY model ($h=0$)}\label{a:XY0}
 In this appendix we detail the diagonalization of the quantum XY model in zero field. 
For the sake of simplicity, we change the parametrization as follows
\be\label{eq:paramXY}
\gamma=\tanh\alpha\qquad J=2\cosh\alpha \qquad (h=0)\, .
\ee
The eigenvalue problem \eref{eq:eig_prob} can be cast in the following linear recurrence equation
\be\label{eq:rec}
\left\{
\begin{array}{l}
\sigma^y e^{\alpha \sigma^z}\vec u_2=\varepsilon \vec u_1\\
\sigma^y e^{-\alpha \sigma^z}\vec u_{n-1}+\sigma^y e^{\alpha \sigma^z}\vec u_{n+1}=\varepsilon \vec u_n\qquad 1<n<L\\
\sigma^y e^{-\alpha \sigma^z}\vec u_{L-1}=\varepsilon \vec u_L\, ,
\end{array}
\right.
\ee
where, for any given excitation energy $\varepsilon$, we defined
\begin{equation*}
\vec u_i\propto\left(\begin{array}{c}
[\vec{\mathcal U}]_{2i-1}\\
{}[\vec{\mathcal U}]_{2i}
\end{array}\right)\, .
\end{equation*}
We solve the recurrence equation in two steps, reducing it first to a simpler equation.
To this aim we define $\vec v_n$  of the form
\be\label{eq:rec1}
\vec v_n=\vec u_n-e^{i\phi/2} \sigma^y e^{i\beta \sigma^z} \vec u_{n-1}\, ,
\ee
with $\phi$ and $\beta$ two auxiliary parameters.
From the second equation(s) of \eref{eq:rec} we obtain
\begin{equation*}
\fl\qquad e^{\alpha \sigma^z}\vec v_{n+1}=\Bigl[\varepsilon \sigma^y-e^{i\phi/2} \sigma^y e^{(i\beta-\alpha) \sigma^z} \Bigr]\vec v_{n}+ \Bigl[\varepsilon e^{i\phi/2} e^{i\beta \sigma^z}  -e^{i\phi} e^{\alpha \sigma^z}-e^{-\alpha \sigma^z} \Bigr]  \vec u_{n-1}\, ,
\end{equation*}
so it is convenient to impose the condition
\begin{equation*}
\varepsilon e^{i\phi/2} e^{i\beta \sigma^z}  -e^{i\phi} e^{\alpha \sigma^z}-e^{-\alpha \sigma^z}=0\, .
\end{equation*}
This system of two equations has the following solution:
\be\label{eq:lambda}
\left\{
\begin{array}{l}
e^{i\beta}=2\cosh\bigl(\alpha+i\frac{\phi}{2}\bigr)\\
\varepsilon^2=4\cosh\bigl(\alpha+i\frac{\phi}{2}\bigr)\cosh\bigl(\alpha-i\frac{\phi}{2}\bigr)=4(\cos^2\frac{\phi}{2}+\sinh^2\alpha)
\end{array}
\right.
\ee
It is worth remarking that we can already identify the dispersion relation:
\be\label{eq:energy}
\varepsilon(\phi)=2\sqrt{\cos^2\frac{\phi}{2}+\sinh^2\alpha}\, ,
\ee
where however $\phi$ is still an undefined variable. 
Imposing \eref{eq:lambda}, the recurrence equation for $\vec v$ becomes
\begin{equation*}
\vec v_{n+1}=e^{-i\phi/2}\sigma^y e^{(2\alpha+i\beta)\sigma^z}\vec v_{n}\, .
\end{equation*}
This is readily solved
\begin{equation*}
\vec v_{n}=\left\{ \begin{array}{ll}
e^{i(1-n)\phi/2}\sigma^y e^{(2\alpha+i\beta)\sigma^z}\vec u_1&n\ {\rm even}\\
e^{i(1-n)\phi/2}\vec u_1&n\ {\rm odd}\, ,
\end{array}\right.
\end{equation*}
where we used the first equation of \eref{eq:rec} to express the solution in terms of $\vec u_1$. We point out that with periodic boundary conditions one can take the solution $\vec v_n=0$. 

We can go back to the original variables using the following identity
\begin{equation*}
\vec u_{n}=[e^{i\phi/2} \sigma^y e^{i\beta \sigma^z}]^{n-1} \vec u_1+\sum_{j=0}^{n-2}[e^{i\phi/2} \sigma^y e^{i\beta \sigma^z}]^j\vec v_{n-j}\, ,
\end{equation*}
which readily follows from \eref{eq:rec1}. 
We finally obtain ($\phi\neq 0,\pi$)
\begin{eqnarray}\label{eq:evec}
\vec u_{2n-1}&=&\frac{\sin(n\phi) +\sin((n-1) \phi)e^{2\alpha\sigma^z}}{\sin\phi}\vec u_1\nn
\vec u_{2n}&=&\varepsilon\frac{\sin(n\phi)}{\sin\phi}\sigma^ye^{\alpha\sigma^z}\vec u_1\, .
\end{eqnarray}
The admissible values of $\phi$ and, in turn, of the energies, are obtained imposing the last equation of \eref{eq:rec}. 

Before computing the quantization conditions, we note that both eigenvalues and eigenvectors are invariant under $\phi\rightarrow -\phi$ and $\phi\rightarrow \phi+2\pi$, therefore we can restrict ourselves to $\mathrm{Re}[\phi]\in(0,\pi)$ and $\mathrm{Im}[\phi]\geq 0$. 
\subsubsection*{Quantization conditions.}
In the finite chain the `pseudomomentum'  $\phi$ is quantized differently depending on whether the chain is even or odd. 
\paragraph{Even chain.}
For $L$ even, the last equation of \eref{eq:rec} gives
\begin{equation*}
\frac{\sin(\frac{L}{2}\phi) e^{-\alpha \sigma^z}+\sin((\frac{L}{2}-1) \phi)e^{\alpha\sigma^z}}{\sin\phi}\vec u_1=\varepsilon^2\frac{\sin(\frac{L}{2}\phi)}{\sin\phi}e^{\alpha\sigma^z}\vec u_1\, .
\end{equation*}
Using the second equation of \eref{eq:lambda}, this can be recast in the form ($\phi\neq 0,\pi$)
\be\label{a:eq:quantEven}
e^{i(L+1)\phi}\vec u_1=\frac{\cosh(\alpha\sigma^z+i\frac{\phi}{2})}{\cosh(\alpha\sigma^z-i\frac{\phi}{2})}\vec u_1\, .
\ee
Eq.~\eref{eq:quantEven} is satisfied only if $\vec u_1$ is an eigenvector of $\sigma^z$.  
Let us start considering the real solutions. It follows from \eref{eq:quantEven} that the difference between a solution and the next one is $\sim \frac{2\pi}{L+1}$: the left hand side of the equation is invariant under such shift, while the right hand side being modified just by an $O(L^{-1})$ term. 
At the leading order we have
\be\label{eq:quantrealEven}
\fl\qquad\qquad \phi_n\vec u_1=\left[\frac{2\pi n}{L+1}+\frac{2}{L+1}\arctan\Bigl(\tanh\alpha\tan\frac{\pi n}{L+1}\Bigr)\sigma^z\right]\vec u_1+O(L^{-2})\, .
\ee
We point out that for $\sigma^z\vec u_1=-\vec u_1$ the index $n$ can assume any integer value between $1$ and $\frac{L}{2}$, while for $\sigma^z\vec u_1=\vec u_1$ there are solutions only for $n\in \bigl[1,\frac{L}{2}-1\bigr]$. The missed solution is a bound state.  

The bound state can be obtained directly from \eref{eq:quantEven}. Let $\phi=\phi_R+i\eta$, with $\eta$ positive. We have
\be\label{eq:quantboundeven}
e^{i(L+1)(\phi_R+i\eta)}\vec u_1=\frac{\cosh(\alpha\sigma^z+i\frac{\phi_R+i\eta}{2})}{\cosh(\alpha\sigma^z-i\frac{\phi_R+i\eta}{2})}\vec u_1\, .
\ee
In the limit of large $L$, the left hand side of \eref{eq:quantboundeven} approaches zero, therefore the numerator of the right hand side has to zero as $L\rightarrow\infty$. This implies
\begin{equation*}
\Bigl[\alpha\sigma^z+i\frac{\phi_R+i\eta}{2}\Bigr]\vec u_1\rightarrow i\frac{\pi}{2}\vec u_1
\end{equation*}
that is to say 
\begin{eqnarray*}
&\sigma^z\vec u_1=\vec u_1\\
&\phi\rightarrow\pi+2i\alpha\, ,
\end{eqnarray*}
which confirms that the only solution with nonzero imaginary part is in the sector $\sigma^z\vec u_1=\vec u_1$. 
Including also the leading correction we obtain
\begin{eqnarray}\label{eq:BS}
&\mathrm{Re}[\phi]=\pi\nn
&\mathrm{Im}[\phi]\sim 2 \alpha-2\sinh(2\alpha)e^{-2(L+1)\alpha}\, .
\end{eqnarray}
In the following we will use the notation $\mathrm{Im}[\phi]=2\alpha^-$ to emphasize that the imaginary part is close to but smaller than $2\alpha$.

This completes the set of pseudomomenta.

Finally, we report a useful identity
\be\label{eq:transf}
\fl\qquad \varepsilon(\phi)\sin\Bigl(\Bigl(\frac{L}{2}+1-n\Bigr)\phi\Bigr)=-(-1)^{[\phi]}\Bigl[\sin(n\phi)e^{\mp \alpha} +\sin((n-1) \phi)e^{\pm \alpha}\Bigr]
\ee
where the upper (lower) sign applies to the case $\sigma^z \vec u_1=\vec u_1$ ($\sigma^z \vec u_1=-\vec u_1$) and we indicated with $[\phi]$ the position of  $\phi$ in the sequence obtained sorting the solutions according to their real part, from the smallest to the greatest. This identity can be easily proved using the quantization rule \eref{eq:quantEven} and the fact that the difference between two consecutive solutions is $\sim \frac{2\pi}{L+1}$. 

\paragraph{Odd chain.}
If $L$ is odd, the last equation of \eref{eq:rec} gives
\begin{equation*}
\varepsilon \frac{\sin(\frac{L+1}{2}\phi) }{\sin\phi}\vec u_1=0\, .
\end{equation*}
There are two possibilities: either
\be\label{eq:QCodd}
\phi_n=\frac{2\pi n}{L+1}\qquad n=1,\dots,\frac{L-1}{2}\, .
\ee
or $\varepsilon=0$, \emph{i.e.}
\be
\phi=\pi+2i\alpha
\ee
Differently from the even case, the quantization conditions are very simple and do not depend on $\vec u_1$. 

\paragraph{Orthogonality and completeness relations.}
Here we report the orthogonality and completeness relations that follow from \eref{eq:evec}.

\paragraph{Even chain.}
Let $\sigma^z\vec u_1=s \vec u_1$, with $s=\pm 1$, and $\ell,n\in[1,L/2]$. We find
\begin{eqnarray}\label{eq:ortE1}
\fl\ &\sum_{\phi_{(s)}}\frac{\sin(\ell \phi)\sin(n \phi)}{L+1-2s\frac{\sinh(2\alpha)}{\varepsilon^2(\phi)}}=\frac{1}{4}\delta_{\ell n}\\
\label{eq:ortE2}
\fl\ &\sum_{\phi_{(s)}}\frac{[\sin(\ell \phi)e^{-s\alpha}+\sin((\ell-1)\phi)e^{s \alpha}][\sin(n \phi)e^{-s\alpha}+\sin((n-1)\phi)e^{s\alpha}]}{(L+1)\varepsilon^2(\phi)- 2s\sinh(2\alpha)}=\frac{1}{4}\delta_{\ell n}\, .
\end{eqnarray}
\begin{eqnarray}
\fl\ &\sum_{n=1}^{L/2}\frac{\sin(n\phi)\sin(n  \phi')}{L+1- 2s\frac{\sinh(2\alpha)}{\varepsilon^2(\phi)}}=\frac{1}{4}\delta_{\phi,\phi'}\\
\fl\ &\sum_{n=1}^{L/2}\frac{[\sin(\ell \phi)e^{\mp\alpha}+\sin((\ell-1)\phi)e^{s\alpha}][\sin(n \phi')e^{-s\alpha}+\sin((n-1)\phi')e^{s\alpha}]}{(L+1)\varepsilon^2(\phi)- 2s\sinh(2\alpha)}=\frac{1}{4}\delta_{\phi \phi'}
\end{eqnarray}

\paragraph{Odd chain.}
For the real solutions $\phi\in{\mathbb R}$ we have ($\ell,n\in[1,L]$)
\be
\sum_{\phi}\frac{\sin(\ell\phi)\sin(n\phi)}{L+1}=\frac{1}{4}(\delta_{\ell n}-\delta_{\ell, L+1-n})
\ee
\be
\sum_{n=1}^{(L\pm 1)/2} \frac{\sin(n\phi)\sin(n\phi')}{L+1}=\frac{1}{4}\delta_{\phi,\phi'}\, .
\ee

\subsection[\hspace{1.5cm}: Transverse-field Ising chain]{Transverse-field Ising chain}\label{a:TFIC}
In this appendix we sketch the diagonalization of the transverse-field Ising chain and compute the local conservation laws using the direct approach of \sref{s:lclXY}.
 
\subsubsection*{Diagonalization}
The Hamiltonian of the TFIC is given by 
\be\label{eq:HTFIC}
H=-\frac{J}{2} \sum_{j=1}^{L-1}\sigma_\ell^x\sigma_{\ell+1}^x+h\sigma_\ell^z
\ee
Using representation \eref{eq:Hquad}, we identify $\mathcal H$ as the block-Toeplitz matrix with blocks
\be\label{eq:blockTI}
\fl\quad \left(
\begin{array}{cc}
\mathcal H_{2\ell-1, 2n-1}&\mathcal H_{2\ell-1, 2n}\\
\mathcal H_{2\ell, 2n-1}&\mathcal H_{2\ell, 2n}
\end{array}\right)=J (\delta_{n,\ell+1}\sigma^y \frac{1+\sigma^z}{2}+\delta_{\ell,n+1}\sigma^y \frac{1-\sigma^z}{2}-h\sigma^y)\, .
\ee
From now on we set $J=1$. The recurrence equation associated with the eigenvalue problem reads (we multiplied by $\sigma^y$ from the left)
\be\label{eq:recTFIC}
\fl\qquad\qquad\left\{
\begin{array}{l}
 \frac{1+\sigma^z}{2}\vec u_2=(\varepsilon \sigma^y +h)\vec u_1\\
\frac{1-\sigma^z}{2}\vec u_{n-1}+\frac{1+\sigma^z}{2}\vec u_{n+1}=(\varepsilon \sigma^y+h) \vec u_n\qquad 1<n<L\\
\frac{1-\sigma^z}{2}\vec u_{L-1}=(\varepsilon \sigma^y + h) \vec u_L\, ,
\end{array}
\right.
\ee
The first equation is already sufficient to determine $\vec u_1$ (up to a multiplicative constant):
\begin{equation*}
\frac{1-\sigma^z}{2}(\varepsilon\sigma^y+h)\vec u_1=0\Rightarrow \vec u_1=\left(\begin{array}{c}
h\\
-i\varepsilon
\end{array}
\right)\, .
\end{equation*}
Multiplying the second equation of \eref{eq:recTFIC} by $\frac{1\pm\sigma^z}{2}$ from the left we find
\be\label{eq:recsimpleTFIC}
\fl\qquad \vec u_{n+1}=\Bigl[\frac{1-\sigma^z}{2}\frac{1- \varepsilon^2 -h\varepsilon\sigma^y}{h}+\frac{1+\sigma^z}{2} (\varepsilon \sigma^y+h)\Bigr] \vec u_n \qquad n\geq 1\, .
\ee
It is convenient to parametrize the matrix in square brackets as follows
\begin{equation*}
e^{-i k \sigma^x e^{\beta\sigma^y}}\equiv \frac{1-\sigma^z}{2}\frac{1- \varepsilon^2 -h\varepsilon\sigma^y}{h}+\frac{1+\sigma^z}{2} (\varepsilon \sigma^y+h)
\end{equation*}
where
\begin{eqnarray*}
\varepsilon^2=1+h^2-2h\cos k\\
e^{\beta\sigma^y}=\frac{\varepsilon+(h-\cos k)\sigma^y}{\sin k}\, .
\end{eqnarray*}
The reader familiar with the quantum Ising model can already recognize the dispersion relation of the model.  
The solution of \eref{eq:recsimpleTFIC} is straightforward 
\be\label{eq:vectTFIC}
\vec u_n=e^{-i(n-1) k \sigma^xe^{\beta\sigma^y}}\vec u_1=e^{-i n k \sigma^xe^{\beta\sigma^y}}\left(\begin{array}{c}
1\\
0
\end{array}
\right)\, .
\ee
The last equation of \eref{eq:recTFIC} implies
\begin{equation*}
\left(
\begin{array}{cc}
\varepsilon^2-h^2&\frac{i\varepsilon}{h}(1+h^2-\varepsilon^2)\\
0&0
\end{array}\right)e^{-i (L-1) k \sigma^x e^{\beta\sigma^y}}\left(\begin{array}{c}
1\\
0
\end{array}
\right)=0
\end{equation*}
that is to say
\begin{equation*}
\left(\begin{array}{c}
i\varepsilon\\
h
\end{array}
\right)\propto e^{-i L k \sigma^x e^{\beta\sigma^y}}\left(\begin{array}{c}
1\\
0
\end{array}
\right)
\end{equation*}
This equation results in the constraint
\be
e^{2i (L+1) k}=\frac{e^{i k}-h}{e^{-i k}-h}\, .
\ee
\subsubsection*{Orthogonality and completeness relations.}
The relations of orthogonality and completeness that follow from \eref{eq:vectTFIC} are the following:
\be
\sum_k \frac{[h\sin (n k)-\sin((n-1)k)][h\sin (\ell k)-\sin((\ell-1)k)]}{\varepsilon^2(k) L+h(h-\cos k)}=\frac{1}{2}\delta_{\ell n}
\ee
\be\label{eq:ortrTFIC}
\sum_k \frac{\varepsilon^2(k) \sin (n k)\sin(\ell k)}{\varepsilon^2(k) L+h(h-\cos k)}=\frac{1}{2}\delta_{\ell n}
\ee 
\be
\sum_{n=1}^L \frac{[h\sin (n k)-\sin((n-1)k)][h\sin (n p)-\sin((n-1)p)]}{\sqrt{ \varepsilon^2(k) L+h(h-\cos k)}\sqrt{\varepsilon^2(p) L+h(h-\cos p)}}=\frac{1}{2}\delta_{k p}
\ee
\be
\sum_{n=1}^L \frac{\varepsilon(k)\varepsilon(p) \sin (n k)\sin (n p)}{\sqrt{\varepsilon^2(k) L+h(h-\cos k)}\sqrt{\varepsilon^2(p) L+h(h-\cos p)}}=\frac{1}{2}\delta_{k p}\, .
\ee

\subsubsection*{Excitations.}
The excitations are obtained from the eigenvectors of $\vec u$ as shown in \eref{eq:bU}. We obtain
\be
b^\dag_k=\sum_{n=1}^L \frac{[h\sin (n k)-\sin((n-1)k)]a_n^x- i \varepsilon(k) \sin (n k)a_n^y}{\sqrt{2 \varepsilon^2(k) L+2h(h-\cos k)}}
\ee
For $h<1$ there is a bound state $k=i\eta$ with
\be
e^{-2(L+1)\eta}=\frac{e^{-\eta}-h}{e^{\eta}-h}\, .
\ee
At the lowest order of perturbation theory we find
\be
\eta\sim - \log h+(h-h^{-1}) h^{2L+1}\, .
\ee
The excitation energy approaches zero exponentially fast in the system length. 
\subsubsection*{Local conservation laws.}
As in the quantum XY model, the bound state produces three quasilocal conservation laws. 
The local charges have instead the form
\begin{eqnarray}
\fl\qquad\qquad \frac{1}{2}\sum_k \varepsilon(k)\cos(j k) (b^\dag_k-b_k)(b^\dag_k+b_k)\nn
=
-\sum_{n,\ell=1}^L\sum_k \frac{\varepsilon^2(k) \cos(j k) \sin (n k)[h\sin (\ell k)-\sin((\ell-1)k)]}{\varepsilon^2(k) L+h(h-\cos k)}i a_n^y a_\ell^x
\end{eqnarray}
With the help of \eref{eq:ortrTFIC} we can easily identify the symbols~\eref{eq:QTH}
\be
q_T(k)=
\cos(jk)[(\cos k-h)\sigma^y+\sin k\sigma^x]
\ee
\be
q_H(k)=\frac{e^{i(j-1)k}}{2}\sigma^y(e^{i k}-h)\, .
\ee
This is a direct confirmation of the results obtained in \eref{s:lclTFIC} using the general formalism that relies on the algebra of block-Toepliz and block-Toeplitz+Hankel operators. 

\subsection[\hspace{1.5cm}: XY model in a transverse field]{Quantum XY model in a transverse field}\label{a:XYh}
In \sref{s:XY} we obtained the rather unexpected result that the local conservation laws in the quantum XY model in a transverse field have a different structure with respect to those in the transverse-field Ising chain. 
In this appendix we show where these complications come from by sketching the diagonalization of the model. 

The Hamiltonian is given by 
\be\label{eq:HTF}
H=-J \sum_{j=1}^{L-1}\frac{1+\gamma}{4}\sigma_\ell^x\sigma_{\ell+1}^x+\frac{1-\gamma}{4}\sigma_\ell^y\sigma_{\ell+1}^y+\frac{h}{2}\sigma_\ell^z
\ee
The matrix $\mathcal H$ that characterizes the fermion representation \eref{eq:Hquad} is still block-Toeplitz with blocks \eref{eq:blockT}
\be
\fl \left(
\begin{array}{cc}
\mathcal H_{2\ell-1, 2n-1}&\mathcal H_{2\ell-1, 2n}\\
\mathcal H_{2\ell, 2n-1}&\mathcal H_{2\ell, 2n}
\end{array}\right)=\frac{J}{2\cosh\alpha} (\delta_{n,\ell+1}\sigma^y e^{\alpha\sigma^z}+\delta_{\ell,n+1}\sigma^y e^{-\alpha\sigma^z}-2h\cosh\alpha\sigma^y\delta_{\ell n})\, .
\ee
where, as in the quantum XY model, we defined $\alpha$ such that $\gamma=\tanh\alpha$.
Setting again $J=2\cosh\alpha$,  the corresponding recurrence equation reads (\emph{cf.} \eref{eq:rec})
\be\label{eq:recTF}
\fl\qquad\qquad\left\{
\begin{array}{l}
\sigma^y e^{\alpha \sigma^z}\vec u_2=(\varepsilon +2 h\cosh\alpha\sigma^y)\vec u_1\\
\sigma^y e^{-\alpha \sigma^z}\vec u_{n-1}+\sigma^y e^{\alpha \sigma^z}\vec u_{n+1}=(\varepsilon +2 h\cosh\alpha\sigma^y) \vec u_n\qquad 1<n<L\\
\sigma^y e^{-\alpha \sigma^z}\vec u_{L-1}=(\varepsilon +2 h\cosh\alpha\sigma^y) \vec u_L\, ,
\end{array}
\right.
\ee
We point out that, as summarized in \sref{ss:sym}, the magnetic field spoils both  spin-flip symmetry along $x$ and $y$ and the dual symmetry $\gamma\rightarrow \gamma^{-1}$. As a result, we can not obtain the case $\gamma>1$ from $\gamma<1$. 
For $\gamma>1$, $\alpha$ has imaginary part equal to $\frac{\pi}{2}$ and our choice of $J$ is purely imaginary. This, in turn, implies that, for $\gamma>1$,  $\varepsilon$ will be purely imaginary.   

The recurrence equation can be solved in two steps using the parametrization
\be\label{eq:rec1TF}
\vec v_n=\vec u_n-\exp(i\phi\sigma^x e^{\beta\sigma^y}) \vec u_{n-1}\, ,
\ee
where $\phi(\neq 0,\pi)$ and $\beta$ are auxiliary variables.
 Plugging \eref{eq:rec1TF} into \eref{eq:recTF} gives
 \begin{equation*}
 \vec v_{n+1}=e^{-2\alpha\sigma^z}e^{-i\phi \sigma^x e^{\beta\sigma^y}}\vec v_n
 \end{equation*}
with the matrix constraint
\begin{equation*}
e^{-\alpha\sigma^z}\exp(-i\phi \sigma^x e^{\beta \sigma^y}) + e^{\alpha\sigma^z}\exp(i\phi  \sigma^x e^{\beta \sigma^y}) =\varepsilon \sigma^y+2h\cosh\alpha\, .
\end{equation*}
Explicitly this means 
\begin{eqnarray*}
 e^{\beta\sigma^y}=\frac{2\cosh\alpha(\cos\phi-h)\sigma^y-\varepsilon}{2\sinh\alpha\sin\phi}\\
 \varepsilon^2=4(\cosh^2\alpha\ (\cos\phi-h)^2+\sinh^2\alpha\sin^2\phi)\, .
\end{eqnarray*}
The equation for $\vec v_n$ can be readily solved
\begin{equation*}
\vec v_n=[e^{-2\alpha\sigma^z}e^{-i\phi\sigma^x e^{\beta\sigma^y}}]^{n-1}\vec u_1\equiv e^{-i(n-1)\vec \theta\cdot \vec \sigma}\vec u_1\, .
\end{equation*}
where the auxiliary vector $\vec\theta$ is defined in such a way that $e^{i\phi\sigma^x e^{\beta\sigma^y}} e^{2\alpha\sigma^z}=e^{i \vec \theta\cdot\vec\sigma}$.
Going back to the original variables we find
\begin{equation*}
\vec u_n=\sum_{j=0}^{n-1}e^{i (n-1-j)\phi \sigma^x e^{\beta\sigma^y}}e^{-i j\vec \theta\cdot \vec \sigma}\vec u_1\, .
\end{equation*}
This can be solved by expanding $\vec u$ in the eigenstates of $\hat\theta\cdot \vec\sigma$
\begin{equation*}
\vec u_n=\sum_\pm\frac{e^{i (n-1)\phi \sigma^x e^{\beta\sigma^y}}-e^{-i\phi \sigma^x e^{\beta\sigma^y}}e^{\mp in\theta}}{\1-e^{-i\phi \sigma^x e^{\beta\sigma^y}}e^{\mp i\theta}}\frac{\1\pm \hat\theta\cdot\vec\sigma}{2}\vec u_1
\end{equation*}
Using 
\begin{equation*}
\frac{\1}{\1-e^{-i\phi \sigma^x e^{\beta\sigma^y}}e^{\mp i\theta}}=\frac{1}{2}\frac{e^{\pm i\theta}-e^{i\phi \sigma^x e^{\beta\sigma^y}}}{\cos\theta-\cos\phi}=\frac{1}{4}\frac{e^{\pm i\theta}-e^{i\phi \sigma^x e^{\beta\sigma^y}}}{h\cosh^2\alpha- \cos\phi}\, ,
\end{equation*}
we finally obtain
\be
\fl\qquad\vec u_n=\frac{1}{2}\frac{\sinh\alpha \sigma^z }{h\cosh^2\alpha- \cos\phi} \Bigl[e^{-i n\phi \sigma^x e^{-\beta\sigma^y}}-(e^{-\alpha\sigma^z}e^{-i\phi\sigma^x e^{\beta\sigma^y}}e^{-\alpha\sigma^z})^n\Bigr] e^{\alpha\sigma^z}\vec u_1\, .
\ee
The right boundary gives the constraint
\be\label{eq:condmat}
\frac{e^{-i (L+1)\phi \sigma^x e^{-\beta\sigma^y}}-(e^{-\alpha\sigma^z}e^{-i\phi\sigma^x e^{\beta\sigma^y}}e^{-\alpha\sigma^z})^{L+1}}{h \cosh^2\alpha-\cos\phi} e^{\alpha\sigma^z}\vec u_1(\phi)=0\, ,
\ee
where we emphasized that $\vec u_1$ depends on $\phi$. 
This results in the quantization rule
\be
\det\Bigl|\frac{e^{-i (L+1)\phi \sigma^x e^{-\beta\sigma^y}}-(e^{-\alpha\sigma^z}e^{-i\phi\sigma^x e^{\beta\sigma^y}}e^{-\alpha\sigma^z})^{L+1}}{\cos\phi- h \cosh^2\alpha}\Bigr|=0\, .
\ee
Apart from the quantization conditions, which are more involved than in the TFIC or in the quantum XY model with $h=0$, the most worrisome complication is that $\vec u_1$ depends on $\phi$ in a nontrivial way. From the previous equations we see that $\vec u_1(\phi)$ is (the only vector) in the kernel of the matrix appearing in \eref{eq:condmat}. It is reasonable to expect that the dependence on $\phi$ is sufficiently smooth to have a well-defined behavior in the thermodynamic limit. However, the complexity of the problem discourages us from trying to obtain the local conservation laws by ``guessing'' their representation in terms of the occupation numbers, as instead we did in \sref{s:lclXY} and in \ref{a:TFIC}.

\end{document}